\documentclass[preprint2]{aastex}

\shorttitle{BH spin in GRBs}
\shortauthors{Janiuk, Moderski \& Proga}

\begin{document}

\title{ 
On the duration of long GRBs: effects of black hole spin} 

\author{A. Janiuk\altaffilmark{1}, R. Moderski\altaffilmark{1},
  D. Proga\altaffilmark{2}}

\altaffiltext{1} {Copernicus Astronomical Center, Bartycka 18, 00-716
  Warsaw, Poland}
\altaffiltext{2} {University of Nevada, Las Vegas, 4505 Maryland Pkwy,
  NV ~89154, USA}

\begin{abstract}
  In the frame of the collapsar model for long gamma ray bursts (GRBs), 
  we investigate the formation of a torus around
  a spinning BH and we check what rotational properties a progenitor star must have 
  in order to sustain torus accretion over relatively long activity periods. 
  We also
  study the time evolution of the BH spin parameter.
  We take into account the coupling between BH mass, its
  spin parameter and the critical specific angular momentum of accreting gas, needed for the torus
  to form. 
  The large BH spin
  reduces the critical angular momentum
  which in turn can increase the GRB duration with respect to the Schwarzschild 
  BH case. 
  We quantify
  this effect and estimate the GRB durations in three cases: 
  when a hyper accreting
  torus operates or a BH spins very fast or both. We show under what conditions
  a given progenitor star produces a burst that can last as short
  as several seconds and as long as several hundred of seconds.
  Our models indicate that it is possible for a single collapse to
  produce three kinds of  jets: 
  (1) a very short, lasting between a fraction of a second and a few seconds, 
  'precursor' jet, powered only by a hyper accreting torus before the BH spins 
  up, (2) an 'early' jet, lasting several tens of seconds and 
  powered by both hyper accretion and BH rotation, and 
  (3) a 'late' jet, powered only by the spinning BH.

\end{abstract}


\section{Introduction}
The commonly accepted  mechanism for a long gamma
ray burst (GRB) production invokes a collapsar
scenario (Woosley 1995; Paczy\'nski 1998; MacFadyen \& Woosley 1999).
In this model the material from the collapsing star feeds
the accretion disk, then the accretion energy is being transferred to
the jet, which in turn produces
gamma rays at some distance from the  central engine. 
Therefore the whole event cannot last much longer than the existence
of a rotationally supported torus in the collapsar center.
Within the collapsar model the jet can also be produced
by a rotating black hole (BH) which can be spun up by the accreting
torus material.

Among the most plausible mechanisms of the energy extraction from the
accretion flow are the neutrino-antineutrino
annihilation (Mochkovitch et al. 1993), 
or the magnetic fields (e.g. Blandford \& Payne 1982; Contopoulos 1995;
Proga et al. 2003). 
The neutrino cooling
(e.g. Popham, Woosley \& Fryer 1999; Di Matteo et al. 2002; 
Janiuk et al. 2004) 
is effective only if the accretion rate is  large ($\dot m
\gtrsim 0.01 M_{\odot}$ sec$^{-1})$).  Also, a large
BH spin ($A_{\rm Kerr} \gtrsim 0.9$)  is
thought to be a necessary condition for the jet launching:
for $A_{\rm Kerr}\sim 0.9$, about 1\% of the accreted rest-mass energy is 
emitted back as a Poynting jet 
(Blandford \& Znajek 1977; McKinney 2005).

On the other hand, the rotationally supported torus may form only when
the substantial amount of specific angular momentum is carried in the
material.  In our recent article (Janiuk \& Proga 2008; hereafter
Paper I) we studied the problem of whether the collapsing star
envelope contained enough specific angular
momentum in order to support the formation of the torus.  This
condition was parametrized by the so called critical specific angular
momentum which in case of a non-rotating BH depends only on its mass. 
In the
present work, we take into account also the 
BH rotation and the coupling between the specific
angular momentum of the accreting material, the BH mass and its spin.

We show that as in Paper I, during the collapse the amount
of the rotating material, which was initially available for the torus
formation, may later become insufficient to support the torus.  
Moreover, the spin of the BH is
changed by accretion (see e.g.  recent studies by 
 Gammie et al. 2004; King \& Pringle 2006; Belczy\'nski et al. 2007).  
In our models, depending on the accretion scenario, both the
spin-up and spin-down of BH are possible, because
part of the infalling material has very small specific angular momentum.

The outline of this paper is as follows.
In Section \ref{sec:model}, we briefly describe the model of the
 evolution of the collapsing star
together with the initial conditions.  
The results are presented in Section \ref{sec:results}; the 
GRB durations are estimated in Section \ref{sec:durations}.  In
Section \ref{sec:diss}, we discuss results in the context of a long GRB
production mechanisms and conditions for the distribution of the
specific angular momentum in a progenitor star.  In
  Appendix \ref{sec:rafal} we provide some formulae for the
  description of the mass and spin evolution of the BH.

\section{Model}
\label{sec:model}
The model is essentially the same as in 
Paper I, with one important modification, namely the
 evolution of the spin of the BH.
As the initial conditions, we use the
spherically symmetric model of the 25.6 $M_{\odot}$ pre-supernova
(Woosley \& Weaver 1995).  The density and mass profiles were shown in
Figure 1 of Paper I.
The mass of the iron core is equal to 1.7 solar masses, 
while the envelope mass,
equal to 23.9 $M_{\odot}$, is available for accretion.

The distribution of the specific angular momentum within a star is
parametrized as either a function of the polar angle $\theta$
(model {\bf A}), or a function of both radius $r$ and $\theta$ (model
{\bf D}). In Paper I, we considered  two more models, i.e., those named
{\bf B} and {\bf C},
but their did not differ significantly from 
models {\bf A} and {\bf D}, respectively. 
Therefore we focus here only on the two models.

In model {\bf A}, we assume the specific angular momentum, $l_{\rm spec}$,
to depend
only on the polar angle:
\begin{equation}
l_{\rm spec} = l_0 f(\theta) \, ,
\end{equation}
where
\begin{equation}
 f(\theta) = 1- |\cos \theta| \, .
\label{eq:ft1}
\end{equation}

The normalization, $l_{0}$, of this dependence is
defined with respect to the critical specific angular momentum
, $l_{\rm crit}$, for the seed BH:
\begin{equation}
l_{\rm crit}(M, A) = {2 G M \over c} \sqrt{2 - A +
  2\sqrt{1-A}} \, ,
\label{eq:lcrit}
\end{equation}
so that $l_0=x l_{\rm crit}$, where $x$ is a free parameter.
In the Eq. 3,
$M$ is the initial BH mass
(iron core mass) and $A \equiv A_{\rm Kerr}$ is its initial dimensionless spin parameter
(see Appendix \ref{sec:rafal}).

In model {\bf D}, we assume that the specific angular
momentum depends on the polar angle, as
well as on the radius in the envelope, as:
\begin{equation}
l_{\rm spec} = l_{0} g(r)f(\theta) \, ,
\end{equation}
where
\begin{equation}
g(r) f(\theta)=\sqrt{r \over r_{\rm in}} \sin^{2}\theta \, ,
\end{equation}
 $r_{\rm in}$ is the inner radius of the envelope, and $l_{0}$ is given below the 
 Eq. \ref{eq:lcrit}. 
%

%
The model {\bf D} corresponds to a constant ratio between the
centrifugal and gravitational forces.  Note that the strong increase
of $l_{\rm spec}$ with radius will lead to a very fast rotation at
large radii.  Therefore, a cut off may be required at some maximum
value, $l_{\rm max}$ (Section \ref{sec:radius}).

The normalization of the models is chosen such that the specific
angular momentum is always equal to the critical value at $\theta =
90^{\circ}$ (and at $r=r_{\rm in}$ if the model depends on radius).
In Section \ref{sec:results}, we present the results of our
calculations considering a range of $x$.

Initially, the mass of the BH is given by the mass of the iron
core of the star, $M = M_{\rm core}$.  The initial conditions for the
torus formation in the collapsar are such that only a fraction of the
envelope mass carries the specific angular momentum larger than the
(initial) critical value.  As shown in Eq. \ref{eq:lcrit} , 
$l_{\rm crit}$ is defined by the mass
of BH, $M$, and its spin, $A_{\rm Kerr}$.  However, as the collapse
proceeds, the mass of BH will increase and also its spin
will change (increase or decrease, depending on the accretion
scenario).  Therefore the critical specific angular momentum will be
changing as well.

To compute the mass of the part of the envelope
 that has specific angular momentum large
enough to form a torus around a given BH, and to estimate the time
duration of the GRB powered by accretion, we need to know the BH mass and spin.

We assume that at each step of the evolution the BH
grows by accreting a mass $\Delta m^{\rm k}$:
\begin{equation}
M^{\rm k} = M^{\rm k-1}+\Delta m^{\rm k} ,
\end{equation} 
and that the BH angular momentum changes as:
\begin{equation}
J^{\rm k} = J^{\rm k-1}+\Delta J^{\rm k} .
\end{equation} 
Here the increment of mass of BH is :
\begin{equation}
\Delta m^{\rm k} = 2 \pi \int_{r_{\rm k}}^{r_{\rm k}+\Delta r_{\rm k}}
\int_{0}^{\pi} \rho(r,\theta) r^{2} \sin \theta d\theta dr ,
\label{eq:deltam}
\end{equation}
and the accreted angular momentum is:
\begin{equation}
\Delta J^{\rm k} = 2 \pi \int_{r_{\rm k}}^{r_{\rm k}+\Delta r_{\rm k}}
\int_{0}^{\pi} \min(l(r,\theta), l_{\rm crit}(M,A)) \rho(r,\theta) r^{2} \sin \theta d\theta dr .
\label{eq:deltaj}
\end{equation}
In the above equation we take into account the fact 
that the angular momentum larger than $l_{\rm crit}$
is not accreted onto the BH, but transported outwards. In this way we provide 
the physical condition for the spin parameter, which must always be $A _{\rm Kerr}\lesssim 1.0$.
 (However, we do not specify any particular mechanism(s) responsible for 
the angular momentum transport.)

The new spin parameter will then be equal
to:
\begin{equation}
A^{\rm k} = {c J^{\rm k} \over G (M^{\rm k})^{2}} .
\end{equation}

\noindent
In the next step of the iteration, both the new parameter and new
mass of BH will affect the critical specific angular
momentum.  Now, depending on the accretion scenario, the part of the
envelope material determined by the new $l_{\rm crit}$, will accrete
onto BH.

We consider here three possible
accretion scenarios:
\begin{enumerate}
\item the accretion onto BH proceeds at the same rate both
  from the torus and from the gas close to the poles (uniform
  accretion);
\item the envelope material with $l<l^{\rm k}_{\rm crit}$ falls on the
  BH first.  Thus, until the polar funnel is evacuated
  completely, only this gas contributes to the BH mass.  After
  that, the material with $l>l^{\rm k}_{\rm crit}$ accretes;
\item the accretion proceeds only through the torus, and only this
  material contributes to the BH growth.  In this case the
  rest of the envelope material is kept aside until the torus is
  accreted.
\end{enumerate}

The above iterative procedure and the accretion scenarios were
described in Paper I and
illustrated in Figure~2 there.
The main modification in the present work is the non-zero spin
parameter of BH, which leads to a different
initial condition for $l_{\rm crit}$ and more complex evolution of the
collapsar. Now $l_{\rm crit}$ is coupled to both
the BH mass $M$ and the spin parameter $A_{\rm Kerr}$.

Due to the increasing BH mass and its changing spin,
the critical angular momentum also changes.  We always stop the calculations,
when there is no material with $l>l^{\rm k}_{\rm crit}$, i.e. able to
form a torus.  However, in a real situation 
the GRB prompt phase will be 
  stopped earlier, i.e. if
  the free fall timescale is too large, 
  or the accretion rate is too small, to be adequate to
  power the GRB ($\dot m = 0.01-1.0$ M$_{\odot}$ s$^{-1}$). 
  Also, in the present model it is important that
  the BH spin parameter is large during the GRB emission.

The duration of the GRB is estimated as the ratio between
 the mass accreted through the torus, and the accretion rate $\dot m$:
\begin{equation}
T_{\rm GRB} = {M_{\rm accr}^{\rm torus} \over <\dot m>},
\label{eq:deltat}
\end{equation}
assuming that the GRB prompt emission is equal to the duration of the
torus replenishment.  The accretion rate, $\dot m$ depends on time and we
determine it instantaneously during the iterations by
the free fall velocity of gas in the torus.
Finally, we impose the conditions for a minimum accretion rate and 
the minimum spin parameter. We then estimate the GRB duration as the ratio between 
the total mass accreted in the torus and the mean accretion rate.

\section{Results}
\label{sec:results}

\subsection{Models with the specific angular momentum dependent on  $\theta$}
\label{sec:theta}

In this Section, we present the results for model {\bf A} of the
specific angular momentum distribution in the collapsing star, 
 and for 
 different accretion
scenarios: (1), (2), and (3). The models are hereafter labeled as 
{\bf A1}, {\bf A2} and {\bf A3}.  
In
this model, $l_{\rm spec}$ does not depend on radius, but only on the
polar angle, $\theta$. The normalization of this distribution,
$x=l_{0}/l_{\rm crit}$, is a
free parameter of our model, and the results are presented as either a
function of $x$, or for some chosen,
exemplary values of $x$.

First, we study how much mass can be accreted onto the BH
during the collapse, both in total and through the rotating torus, as long
as such a torus exists.  This, in the first approximation, will give
an estimate of the GRB duration, because it is proportional to the 
amount of material which is available for accretion.

\begin{figure}
\epsscale{.80}
\plottwo{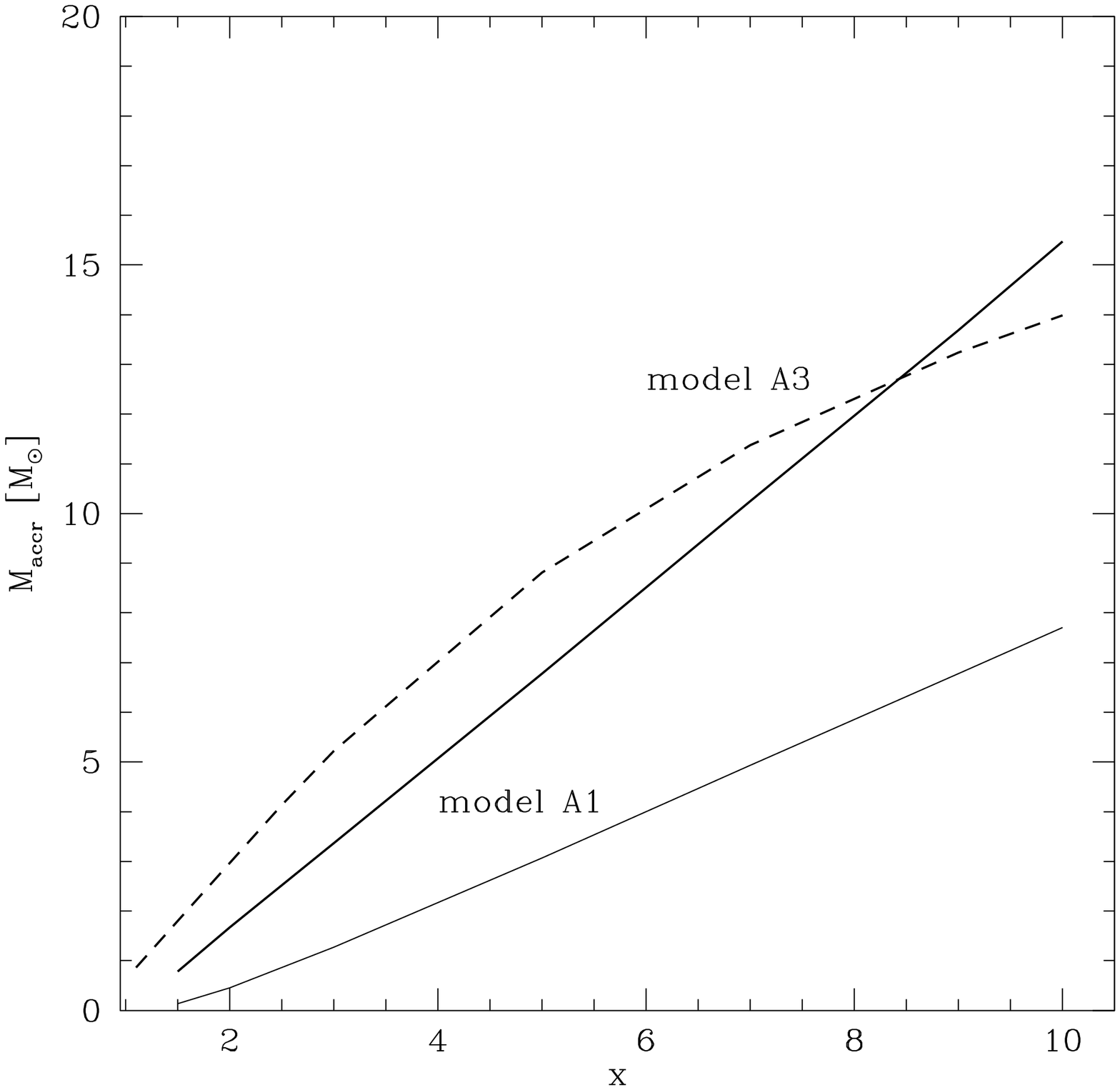}{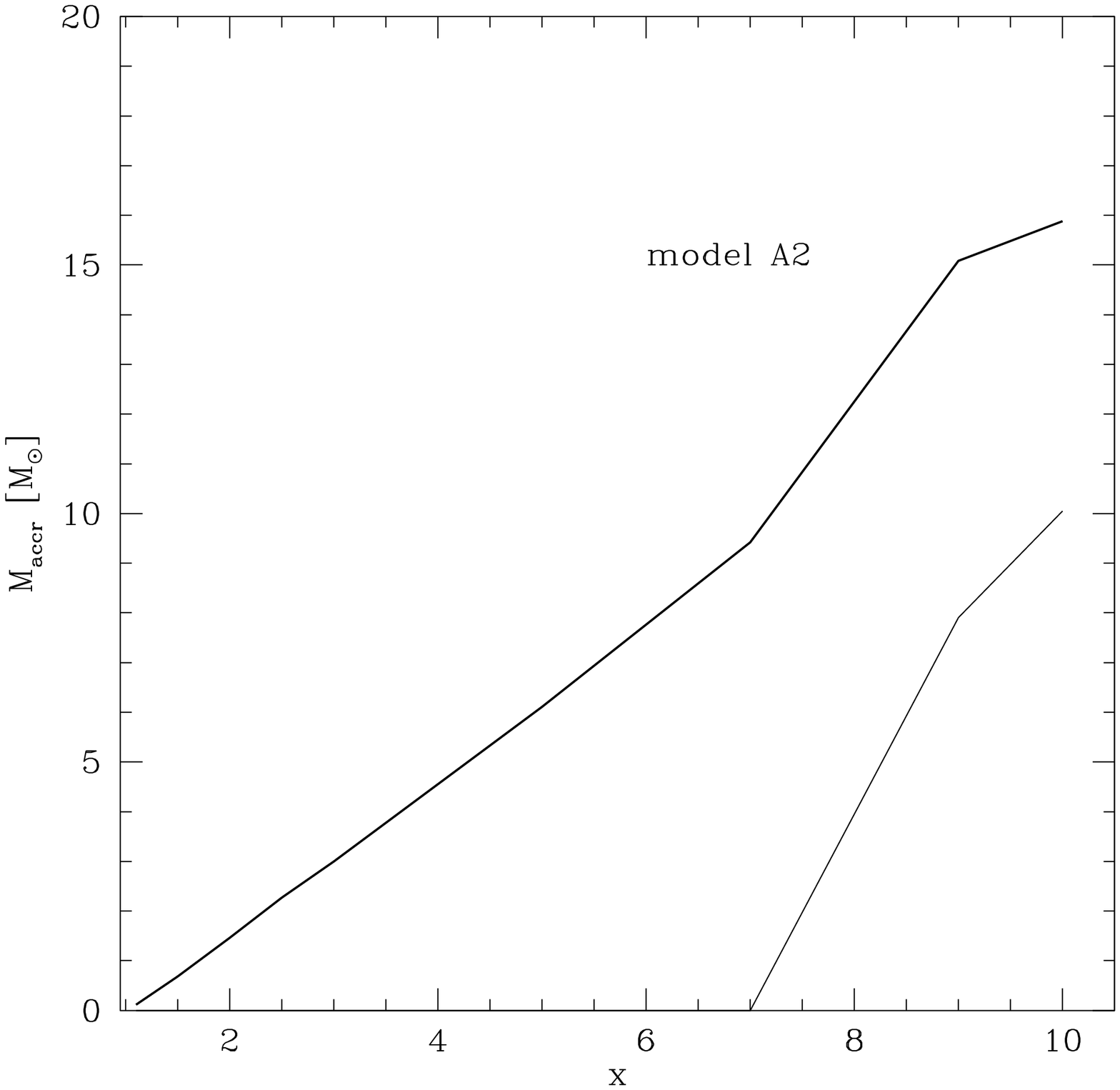}
\caption{The mass accreted onto BH during the collapse, in
  model A for the angular momentum distribution.  Left panel: The
  uniform accretion scenario (A1, solid lines). The two
  lines represent the total accreted mass (thick line) or the mass
  accreted through the torus (thin line). The torus accretion scenario
  (A3) is shown by the dashed line. Right panel: The accretion
  scenario A2, showing the total accreted mass (thick line) and the
  mass accreted through the torus (i.e. in the phase 2; thin line).  }
\label{fig:fig1}
\end{figure}

The Figure \ref{fig:fig1} shows the mass accreted onto the BH, 
as a function of
$x$. Left panel of this Figure shows the scenario {\bf A1}, in which the
mass accretes uniformly.  The thick solid line is
for the total accreted mass, and the thinner line is for this fraction
of mass, which has been accreted through the torus. The single dashed line shows
the scenario {\bf A3}, in which the mass accretes only through the torus.
 We see, that in the model {\bf A3}, the accreted mass can be larger
than in the model {\bf A1} (for $x \le 8$), 
although the accretion in the model {\bf A3} proceeds 
only through the torus. This is because in model {\bf A3}, 
the BH spin can only increase, which in turn lowers the value of $l_{\rm crit}$,
making the condition for $l_{\rm spec}>l_{\rm crit}$ easier to be satisfied. 
Although at the same time the growing BH mass makes the $l_{\rm crit}$ 
increase, for small $x$ this effect is less than the effect of the BH spin. 
In model {\bf A1}, the BH spin decreases (see below), so both the decreasing spin 
and increasing BH mass affect $l_{\rm crit}$ in the same way.

The right panel of the Figure \ref{fig:fig1} shows the scenario {\bf A2}, in which
the matter accretes onto the BH first from the poles, and then
through the torus.  In this scenario, for $x<7$, there is no torus
accretion. This is because the condition for $l_{\rm spec} > l_{\rm crit}$ 
is
never satisfied after the polar material has accreted onto the BH. The total
mass accreted is that from the poles. Only for $x>7$, some fraction of the
envelope material is still capable of forming the torus and accretes
through it (thin line).  This mass adds to the total accreted
mass (thick line).

All the results shown here are
for a rotating BH (the initial Kerr parameter was assumed
$A_{0}=0.85$ and in all the models we had $A_{\rm Kerr}>0$ throughout the collapse; see below).  
In general, the mass accreted onto the spinning BH was 
larger than in the case of the non rotating BH, studied in
Paper I.  For instance, for $x=10$, it is about 15.5 $M_{\odot}$ and 15
$M_{\odot}$ (model {\bf A1}, total accreted mass), 8 $M_{\odot}$ and 7
$M_{\odot}$ (model {\bf A1}, torus accreted mass), and 14 $M_{\odot}$ and 12
$M_{\odot}$ (model {\bf A3}), for a rotating and non-rotating BHs, respectively.

\begin{figure}
\epsscale{.80}
\plottwo{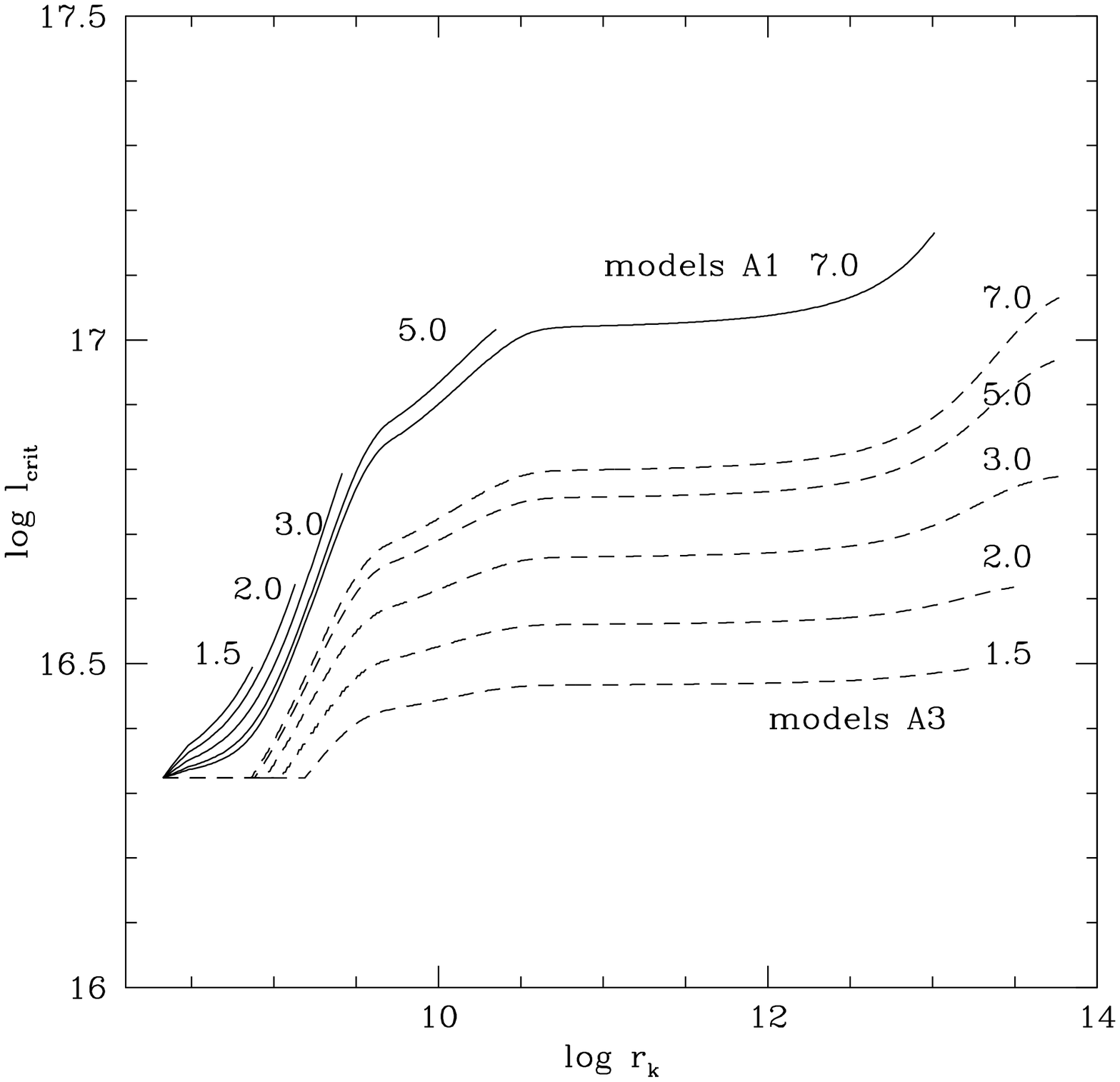}{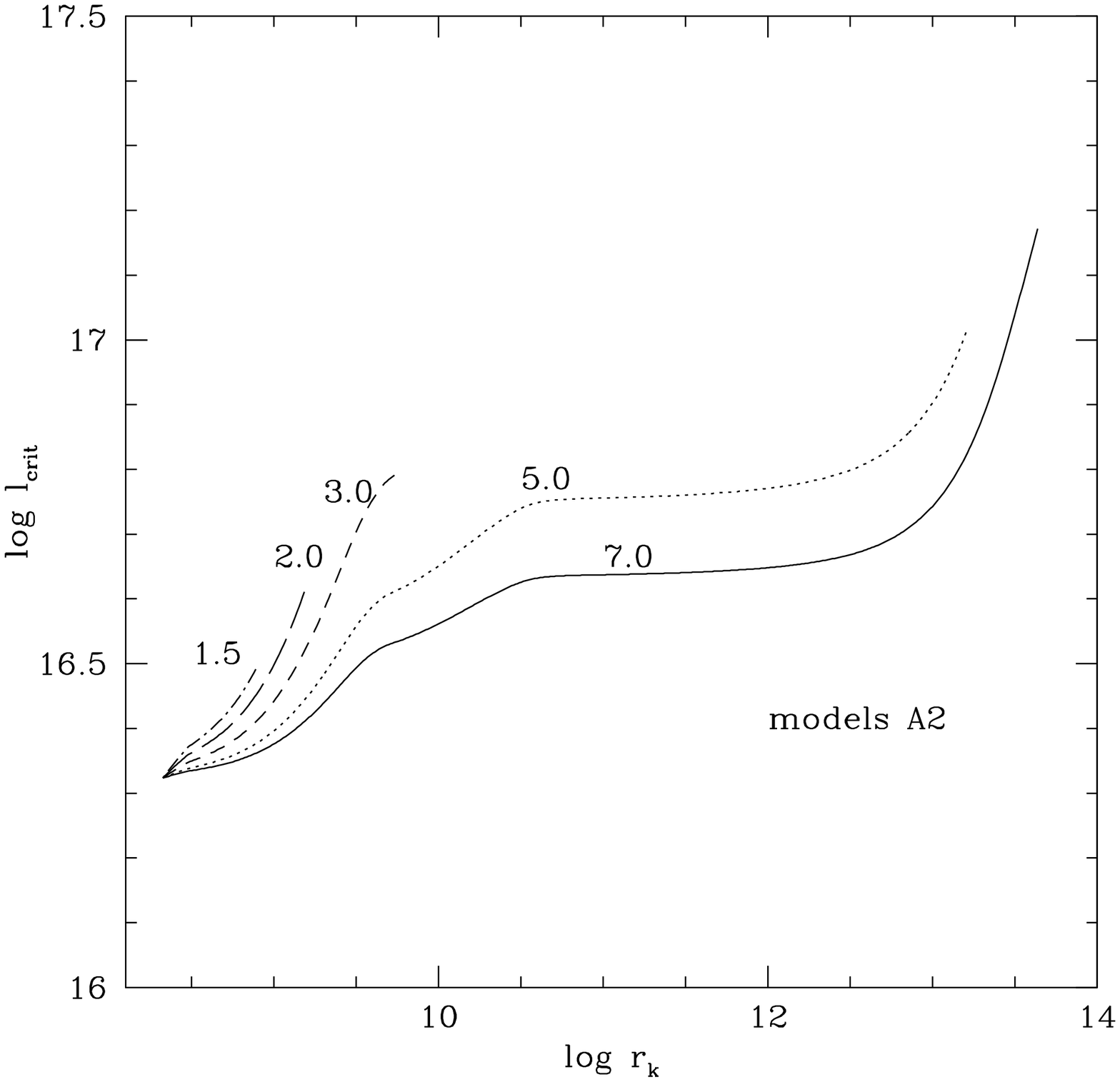}
\caption{The critical specific angular momentum during the collapse, i.e.
  as a function of
  radius $r_{\rm k}$ (the current inner radius of the envelope as it keeps
  accreting onto BH).
  Left panel:
  The solid lines show the uniform accretion scenario (A1), while the
  dashed lines show the torus accretion scenario (model A3), for a
  range of normalizations of the specific angular momentum: x=1.5,
  2.0, 3.0, 5.0 and 7.0, marked on the right for each curve.  Right
  panel: The accretion scenario A2 (only phase 1), for the same
  normalizations $x$.}
\label{fig:figlcrit}
\end{figure}

The above results can be understood, when we study the evolution of the
critical specific angular momentum during the collapse, as shown
in Figure \ref{fig:figlcrit}.  The Figure shows $l_{\rm crit}$ 
as a function of radius.  For the
scenario {\bf A1}, the accretion is uniform and the result  depends only
weakly on $x$. However, for small $x$ the calculations were
stopped earlier, when the torus ceased to exist.
 For scenario {\bf A3}, the
results depend on $x$.
For smaller $x$, the torus mass is smaller, and therefore
less material may accrete onto the BH.
Because the BH is less massive, the increase of $l_{\rm crit}$ 
is slower.
Obviously, the opposite is true in scenario {\bf A2} shown in the
right panel of the Figure \ref{fig:figlcrit} (only the first phase of polar 
accretion is shown, for clarity).

Of course, these results are affected by the large, and changing, Kerr
parameter, $A_{\rm Kerr}$.  Because the term under the square root in the
Eq. \ref{eq:lcrit} is a decreasing function of $A_{\rm Kerr}$, the models with a 
rotating BH always result in a smaller critical angular momentum
than for the non-rotating BH.  Therefore the conditions for the
torus existence in the former models can be satisfied more
easily, and one could expect that the GRB prompt emission can last 
longer than we found in Paper I.

\begin{figure}
\epsscale{.80}
\plottwo{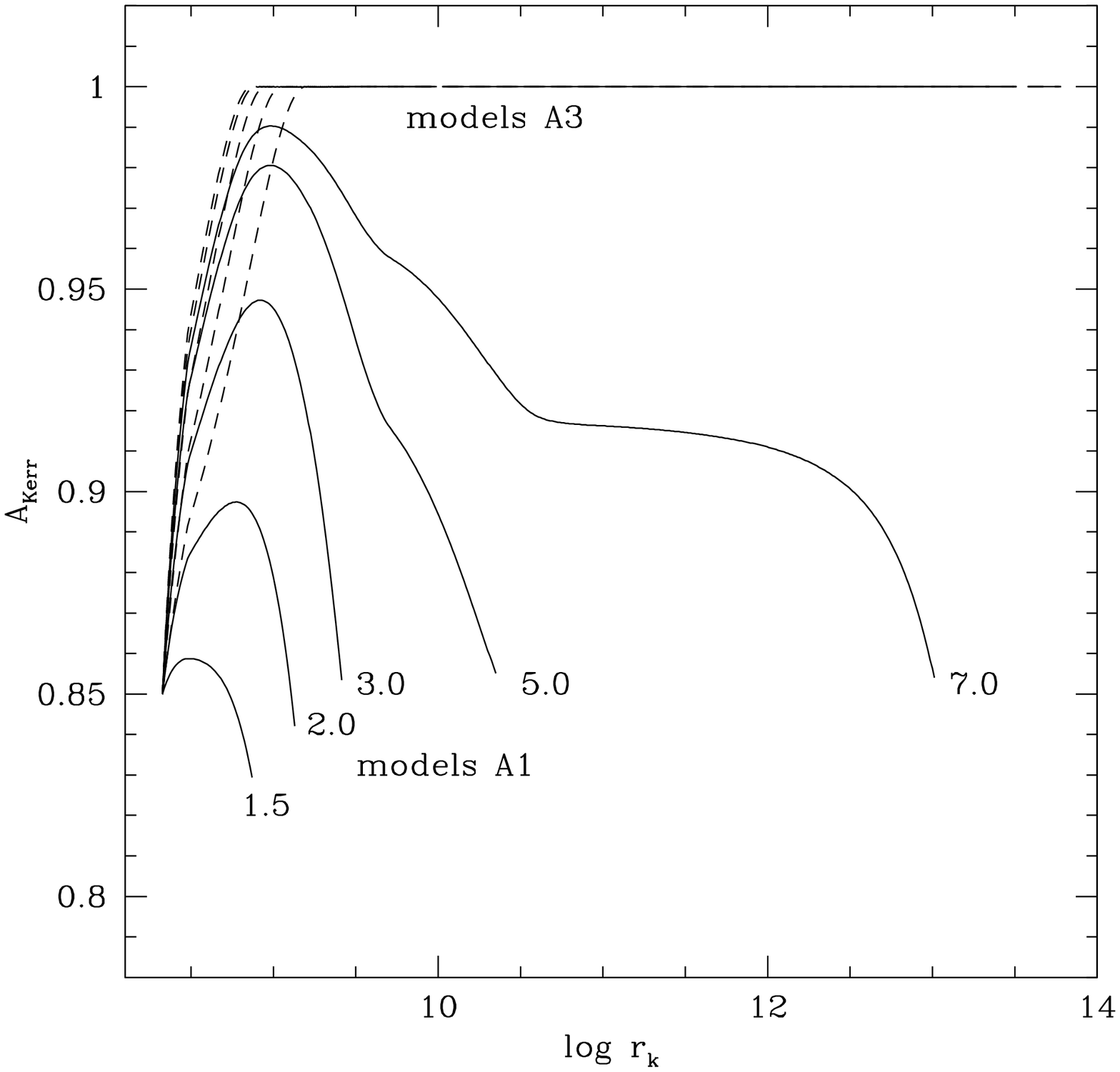}{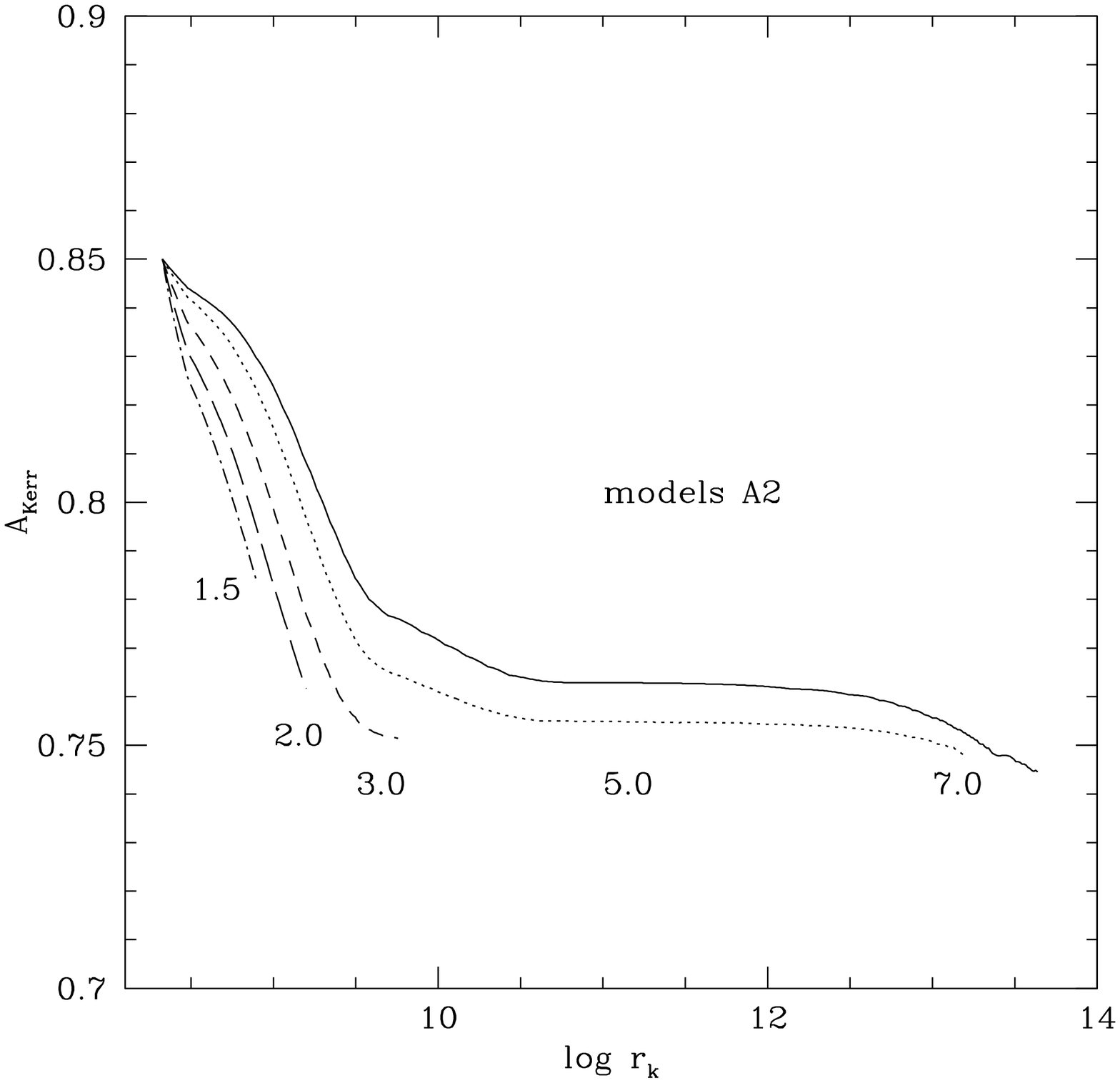}
\caption{The BH spin parameter during the collapse, i.e.
  as a function of
  radius $r_{\rm k}$ (the current inner radius of the envelope as it keeps
  accreting onto BH).
  Left panel: The solid
  lines show the uniform accretion scenario (A1), while the dashed
  lines show the torus accretion scenario (model A3), for a range of
  normalizations of the specific angular momentum: x=1.5, 2.0, 3.0,
  5.0 and 7.0, marked on the right for each curve.  Right panel: The
  accretion scenario A2 (only phase 1), for the same normalizations
  $x$.}
\label{fig:figkerrA_rk}
\end{figure}

The Figure \ref{fig:figkerrA_rk} shows the evolution of $A_{\rm Kerr}$
during the collapse, for scenarios {\bf A1} and {\bf A3} in the left
panel, and for scenario {\bf A2} in the right panel. In the uniform
 accretion scenario {\bf A1}, the BH first spins up, and then spins down. Here, the 
spin evolution depends strongly on $x$.
In scenario  {\bf A2}, BH  spins down during the polar accretion phase, and
the larger $x$, the smaller is the final
spin. 
Then, during the second phase of the model  {\bf A2}, i.e. during the torus accretion,
the BH spins up very quickly, up to  $A_{\rm Kerr}=0.9999$.  The
latter is not shown in the right panel of the Figure \ref{fig:figkerrA_rk} 
for clarity. In
the left panel, the dashed lines mark results for the torus accretion
scenario {\bf A3}, where the final spin is always at $A_{\rm Kerr}=0.9999$ and only 
initially, very weakly, depends on $x$.

We note here that the BH spin never reaches $A_{\rm Kerr}=1.0$ and only approaches 
this value asymptotically. The result of  $A_{\rm Kerr}=1.0$ would be unphysical, 
while the limit 
of $A_{\rm Kerr} \lesssim 1.0$ is naturally provided by our model, in which only the
specific angular momentum of $l_{\rm spec} \le l_{\rm crit}$ contributes to the BH spin.

\begin{figure}
\epsscale{.80}
\plottwo{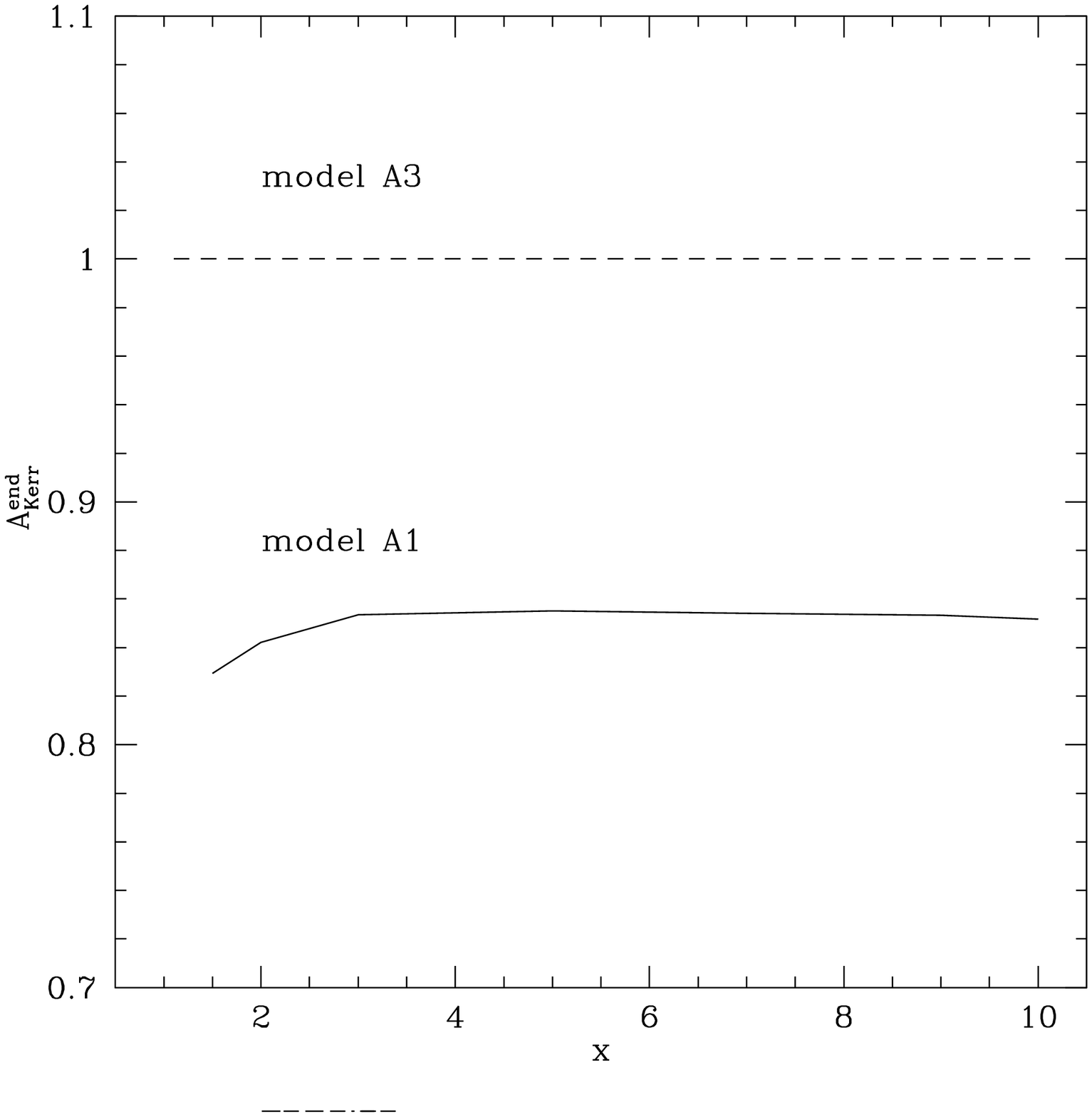}{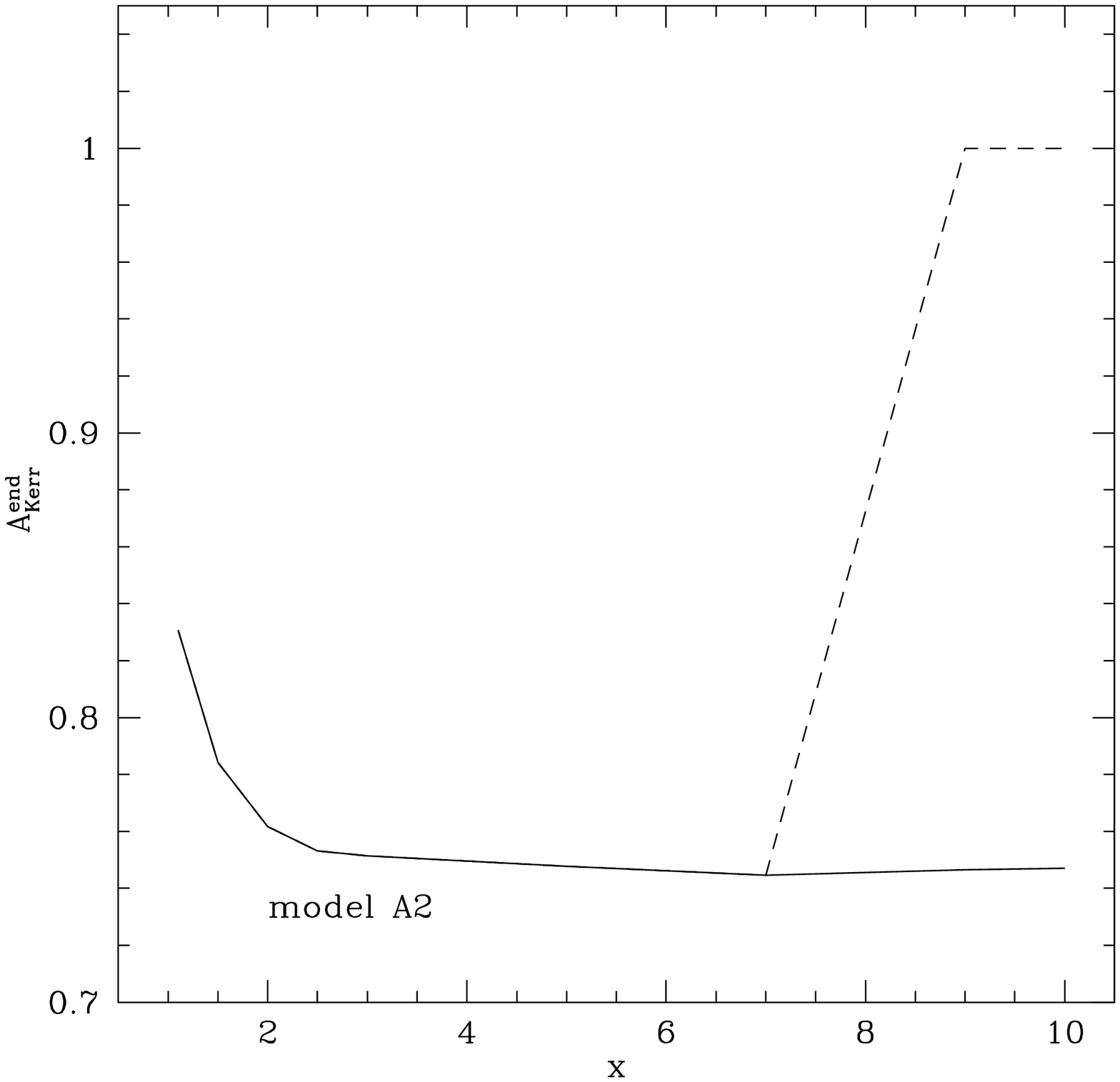}
\caption{The final BH spin parameter after the collapse.  Left
  panel: models of the uniform accretion (A1, solid line), and torus
  accretion (A3, dashed line), as a function of the initial
  normalization of the specific angular momentum.  Right panel: The
  accretion scenario A2, first phase of polar accretion (solid line)
  and second phase of torus accretion (dashed line).}
\label{fig:figkerrA}
\end{figure}

The  evolution of the BH spin is summarized again in the Figure
\ref{fig:figkerrA}.  Here we plot the final value of $A_{\rm end}$, as
a function of $x$. As the Figure shows, $A_{\rm end}$ can be less
than the initial value of $A_{0}=0.85$ for models {\bf A1} and {\bf A2}. 
In other words, the effective spin down of the BH 
is possible either for the uniform accretion, {\bf A1}, but with
a small normalization parameter $x$, or in the two stage
accretion, {\bf A2}, but when the normalization $x$ is so small
that the rotating torus is unable to form.  In other models, the BH
 is either effectively spun-up, to $A_{\rm Kerr}=0.9999$ (model {\bf A3}),
or the final spin does not differ much from the initial one 
(model {\bf A1}, large $x$).

We checked that these results only very weakly depend on the assumed $A_{0}$.
For $A_{0}=0.75$ and  $A_{0}=0.95$, the final 
distribution of
$A_{\rm end}$ with $x$ is also very close to that for $A_{0}=0.85$.
 Interestingly, 
this means that in case of initially rapidly spinning BH with
$A_{0}=0.95$, the object is always effectively spun-down by the uniform accretion.

\subsection{Models with the specific angular momentum dependent on $r$
  and  $\theta$}
\label{sec:radius}

Now, we investigate how the total accreted mass and in consequence the
duration of the GRB will be affected if $l_{\rm spec}$
in the collapsing star is given by a fixed ratio of the centrifugal to
the gravitational force. We discuss here the model {\bf D} for the
$l_{\rm spec}$ distribution, and
the three accretion scenarios are referred to as {\bf D1}, {\bf D2} and {\bf D3}.

\begin{figure}
\epsscale{.80}
\plottwo{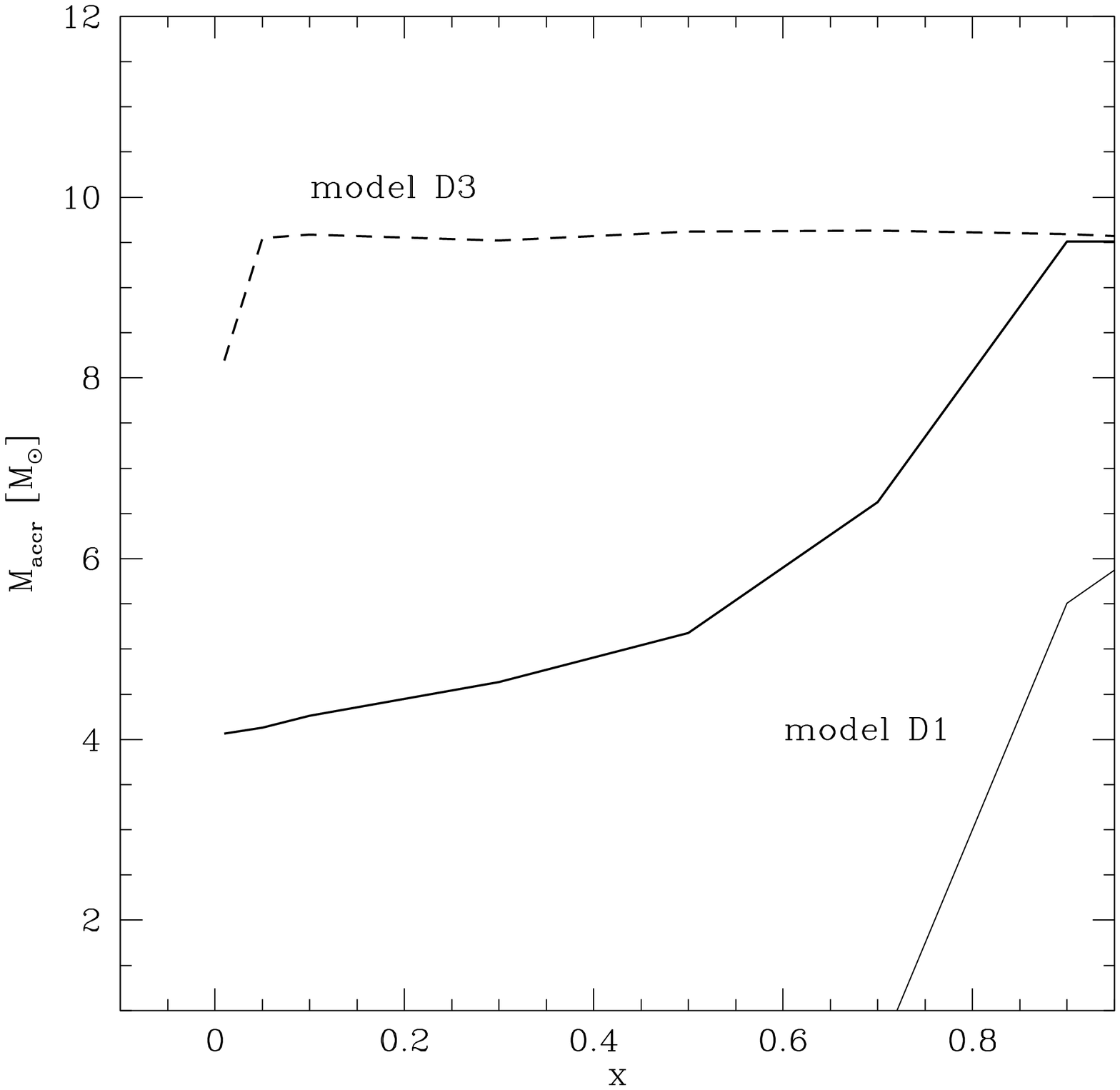}{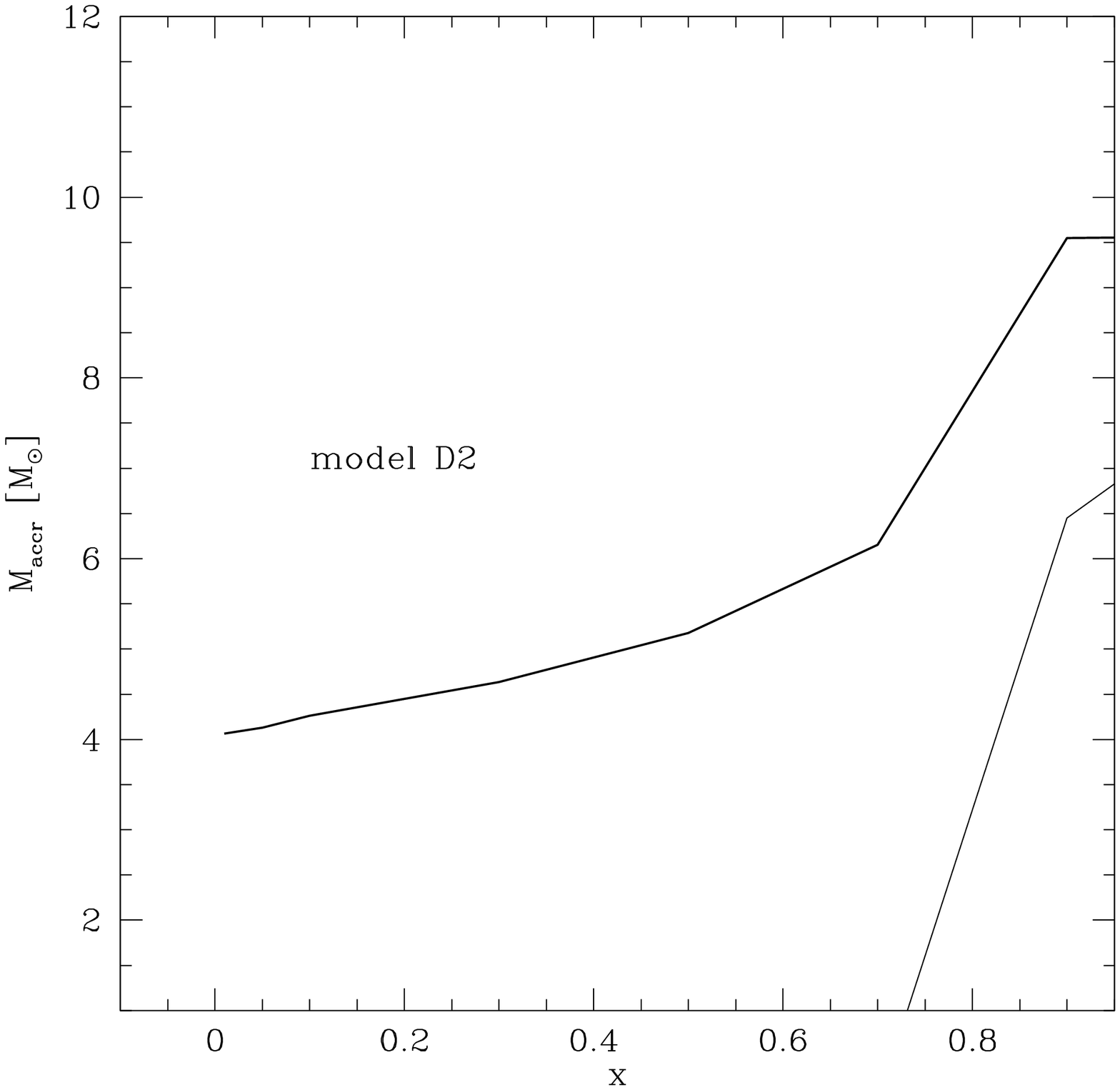}
\caption{The mass accreted onto BH during the collapse, i.e.
  as a function of
  radius $r_{\rm k}$ (the current inner radius of the envelope as it keeps
  accreting onto BH). The plots show 
  model D for the angular momentum distribution.  Left panel: The
  uniform accretion scenario (D1, solid line) is presented by the two
  lines representing the total accreted mass (thick line) and the mass
  accreted through the torus (thin line). The torus accretion scenario
  (D3) is shown by the dashed line. Right panel: The accretion
  scenario D2, showing the total accreted mass (thick line) and the
  mass accreted through the torus (i.e. in the phase 2; thin line).  }
\label{fig:maccrD}
\end{figure}

\begin{figure}
\epsscale{.80}
\plottwo{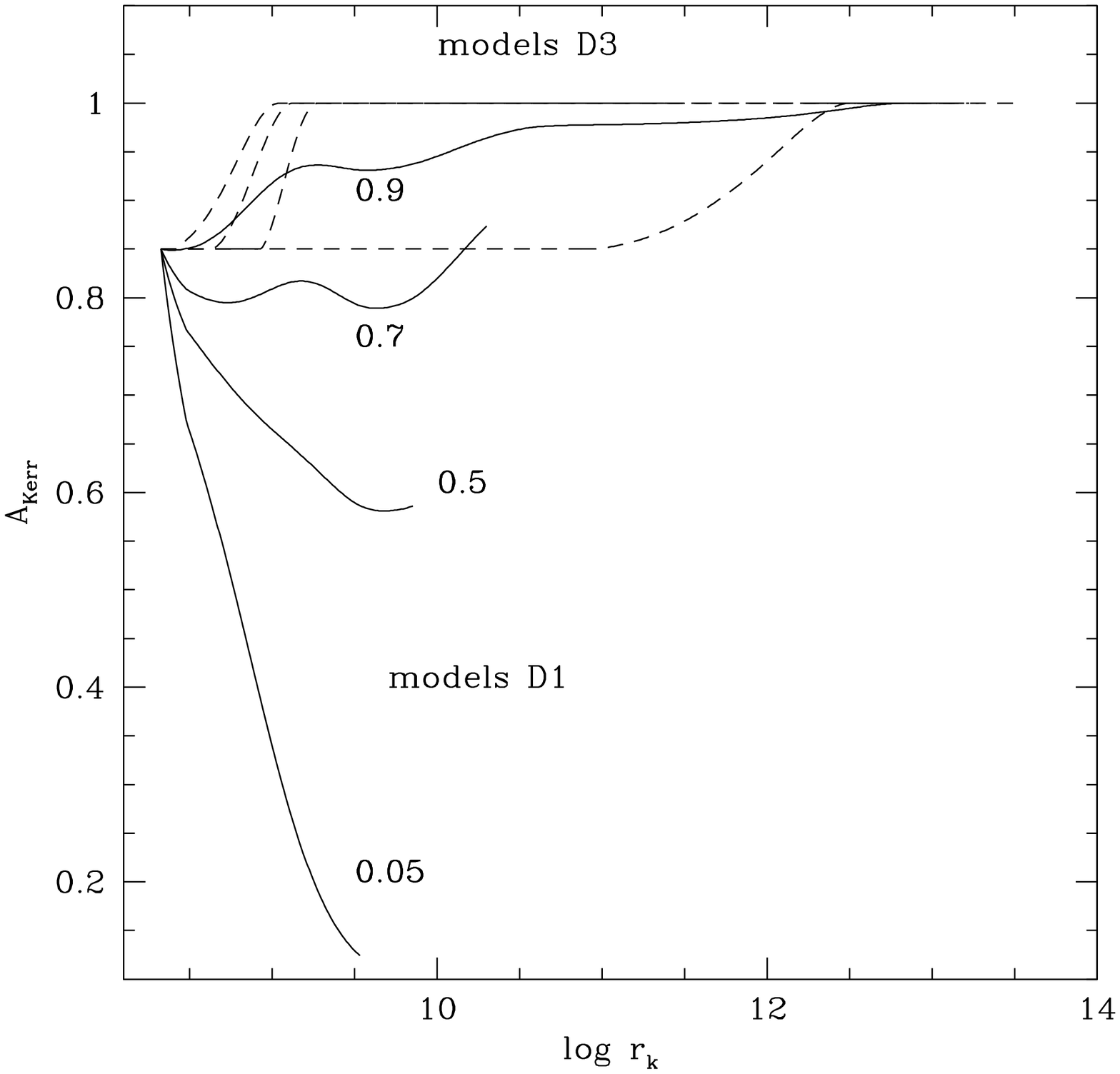}{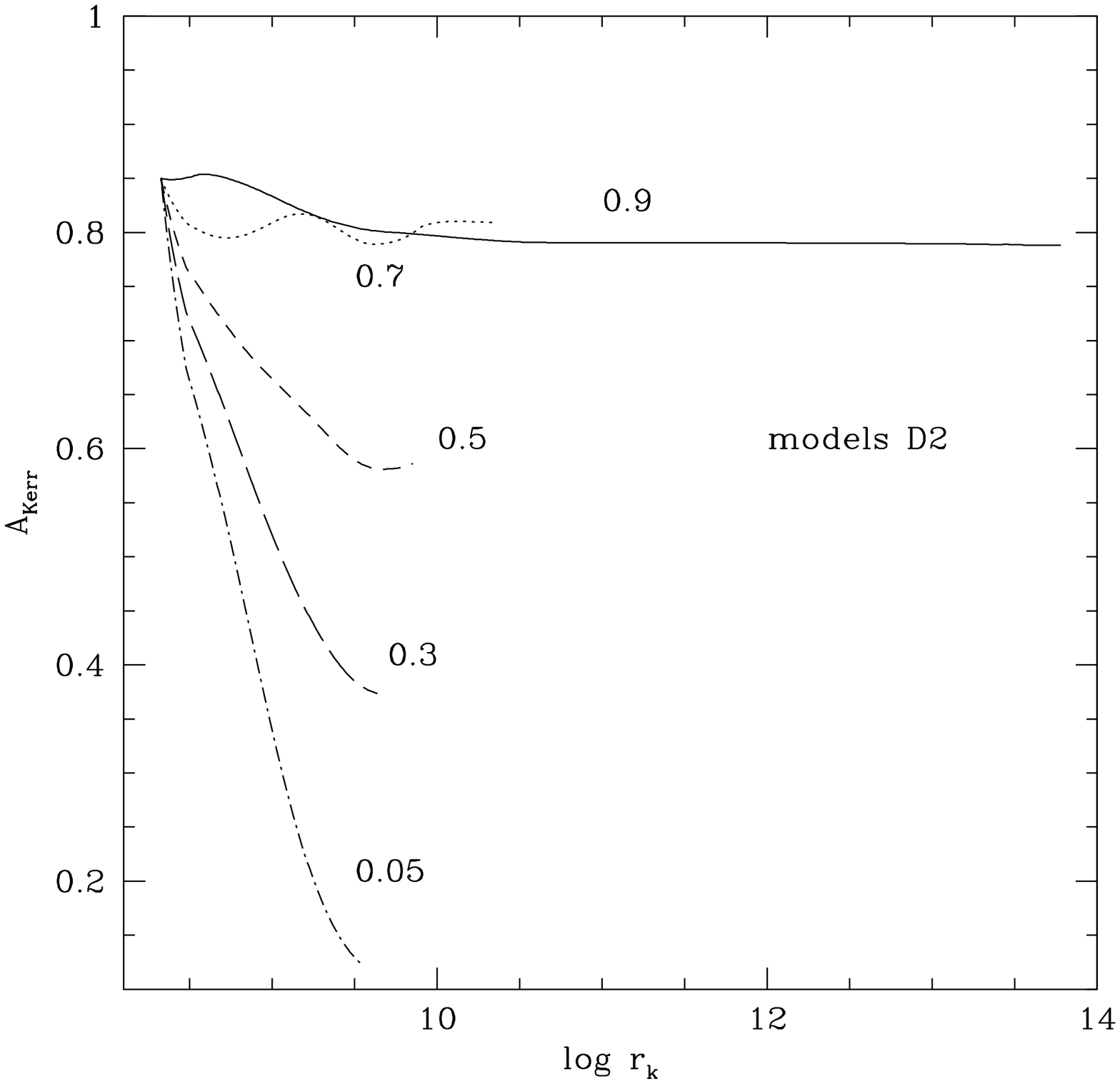}
\caption{The BH spin parameter during the collapse, i.e.
  as a function of
  radius $r_{\rm k}$ (the current inner radius of the envelope as it keeps
  accreting onto BH). 
  Left panel: The solid
  lines show the uniform accretion scenario (D1), while the dashed
  lines show the torus accretion scenario (model D3), for a range of
  normalizations of the specific angular momentum: $x=0.05, 0.5, 0.7$ and 0.9,
  marked on the
  right for each curve.  Right panel: The accretion scenario D2, for
  the same normalizations $x$.}
\label{fig:figkerrD_rk}
\end{figure}

\begin{figure}
\epsscale{.80}
\plottwo{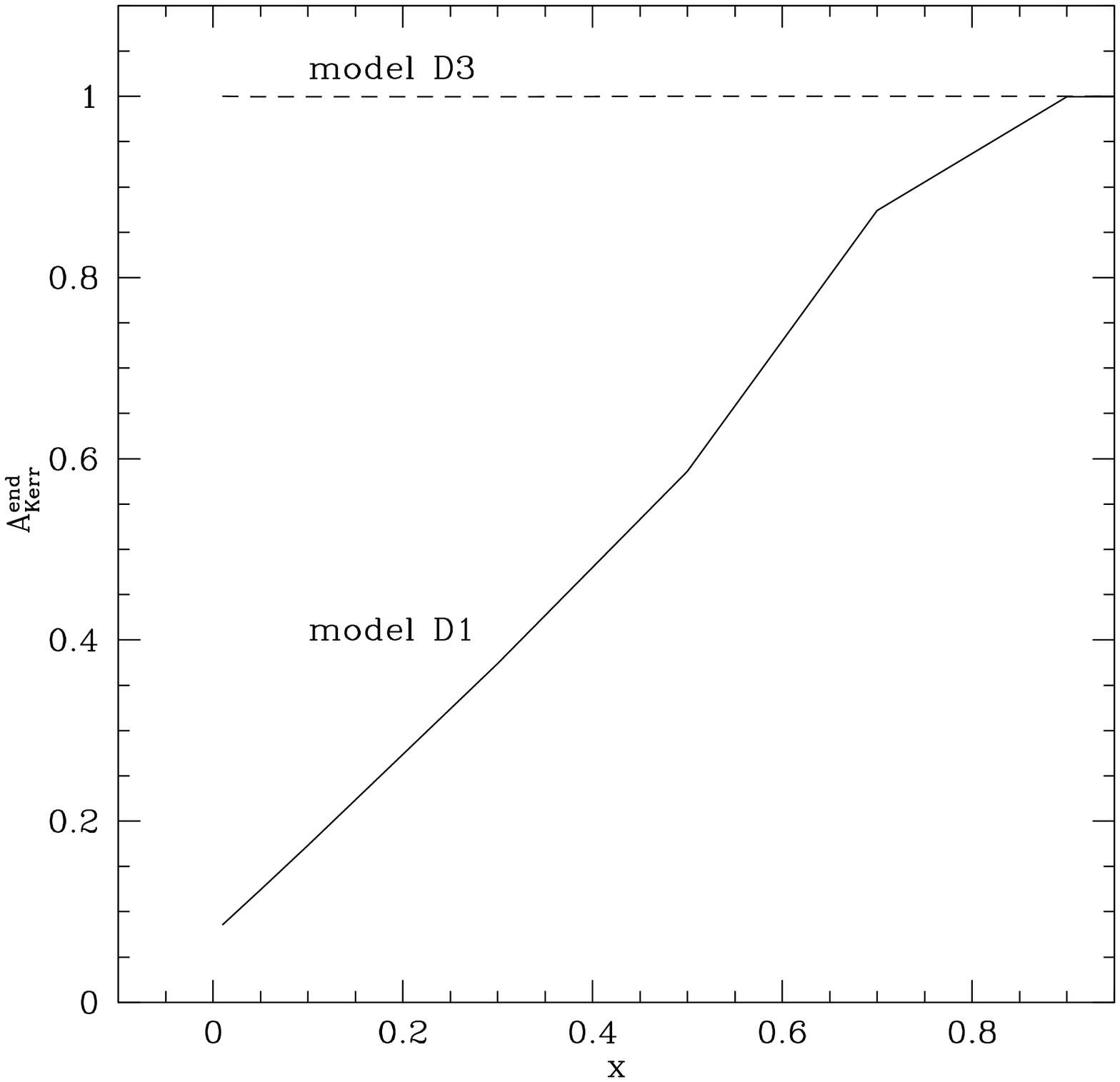}{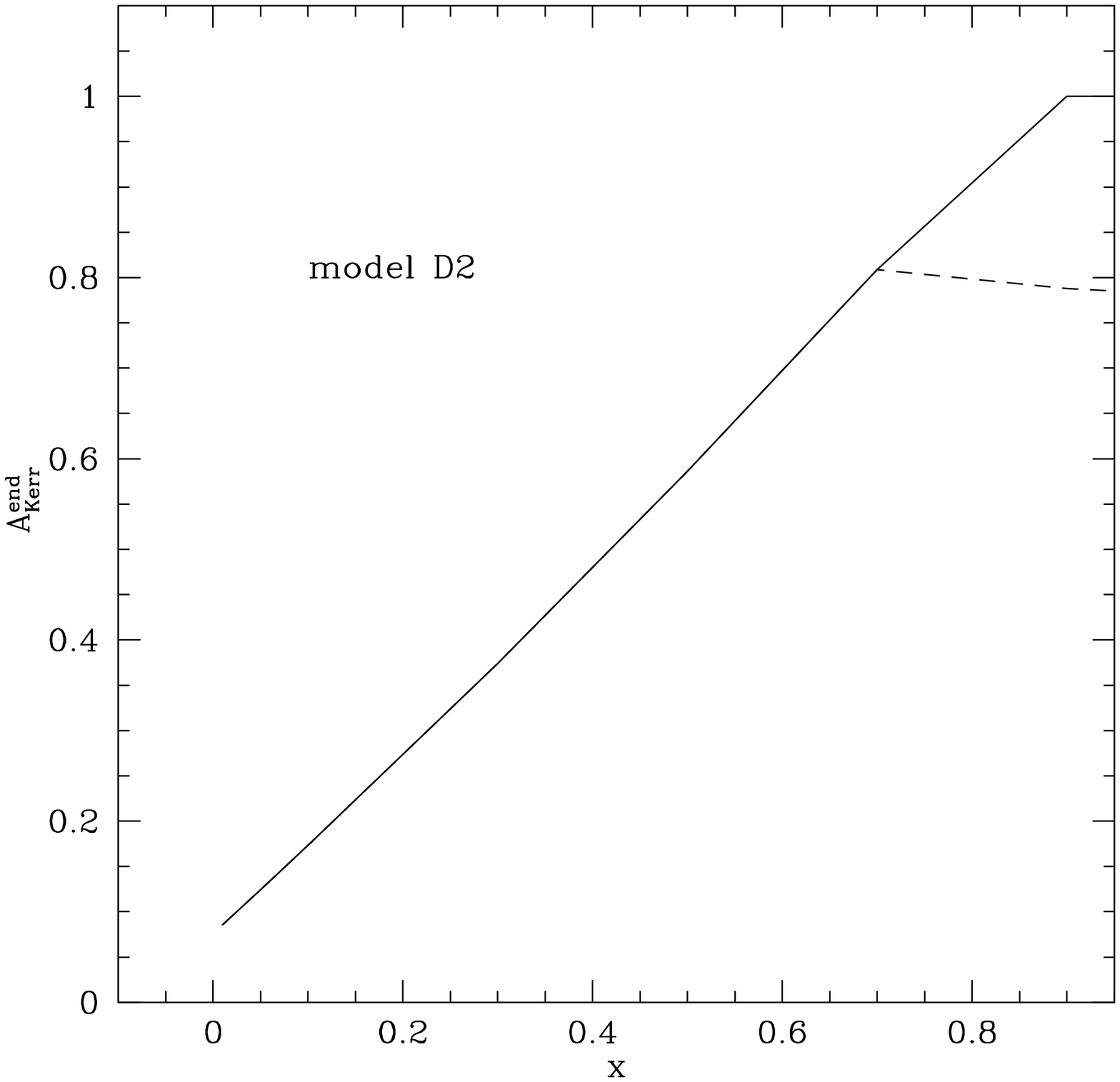}
\caption{The final BH spin parameter after the collapse.  Left
  panel: models of the uniform accretion (D1, solid line), and torus
  accretion (D3, dashed line), as a function of the initial
  normalization of the specific angular momentum.  Right panel: The
  accretion scenario D2, first phase of polar accretion (dashed line)
  and second phase of torus accretion (solid line).}
\label{fig:figkerrD}
\end{figure}

In this model, the specific angular momentum is a strong
function of radius.  Therefore, in a realistic situation
we must have a maximum value of the specific angular momentum, $l_{\rm
  max}$. Here we adopt a moderate value of 
$l_{\rm max}=10^{17}$ cm$^{-2}$s$^{-1}$, following MacFadyen \& Woosley (1999)
and Proga et al. (2003).
The Figure
\ref{fig:maccrD} shows the mass accreted onto the BH in three
accretion scenarios ({\bf D1} and {\bf D3} in the left panel 
and {\bf D2} in the right panel).
Contrary to what was found in Paper I,
the accreted mass
is constant with $x$ only for the torus accretion scenario {\bf D3},
while in models {\bf D1} and {\bf D2} it depends on $x$.
This is because the critical specific angular momentum depends now not only on 
the BH mass but also
on the spin parameter. The BH spin is changing during the collapse in models
{\bf D1} and {\bf D2}, because the
accreting material is not only that from the torus 
(i.e. $l_{\rm spec}>l_{\rm crit}$, which
does not influence the BH spin), but also that from the poles.
The rate of change of the BH spin strongly depends 
on $x$, as we show in Figs. \ref{fig:figkerrD_rk} and \ref{fig:figkerrD} .

The Figure  \ref{fig:figkerrD_rk} shows the BH spin parameter, $A_{\rm Kerr}$,
as a function of radius during the collapse, and the Figure
\ref{fig:figkerrD} shows the final spin $A^{\rm end}_{\rm Kerr}$.  
When the torus accretes, similarly to model {\bf A3}, in the model {\bf D3} the 
BH is spinning up to $A_{\rm Kerr}=0.9999$.
However, in the uniform accretion model {\bf D1}, the BH is effectively spun down for 
most normalizations, i.e. $x<7$. For very small $x$, 
it is even possible for the BH to spin down almost completely at the end of the collapse.
The same is true for the first phase of the scenario {\bf D2}, i.e. the polar accretion.
The existence of a torus in the second phase is possible only
for $x>0.7$ and in this case the BH finally spins up to $A_{\rm Kerr}=0.9999$.
For $x \ge 0.7$ in models D1 and D2, the BH spin slightly fluctuates. This is because 
  of the density profile in the accreting envelope, which is not perfectly smooth, but
  consists of layers, in which various heavy elements are dominant. In models {\bf D},,
the specific angular momentum is a function of radius, which makes the angular momentum accreted onto the BH
much more sensitive to to the position of the current shell, than in case of models {\bf A}.
As a result, for some layers
 the BH may accrete more mass
than the angular momentum, and $A_{\rm Kerr}$ decreases, while for some other 
layers the BH obtains more angular 
momentum than mass, and $A_{\rm Kerr}$ increases (see Eq. \ref{eq:deltam} and \ref{eq:deltaj}).
This is not the case for the model {\bf D3}, 
because here the angular momentum that contributes to the 
BH spin is always given by $l_{\rm crit}$.

\subsection{Duration of a GRB}
\label{sec:durations}

In the first approximation, the duration of a GRB may be proportional to
the mass accreted via the torus during the collapse (as shown in
Figures \ref{fig:fig1} and \ref{fig:maccrD}). However, as the
accretion rate is not constant, the torus accretion will depend also
on the accretion rate (see Eq. \ref{eq:deltat}).

\begin{figure}
\epsscale{.80}
\plottwo{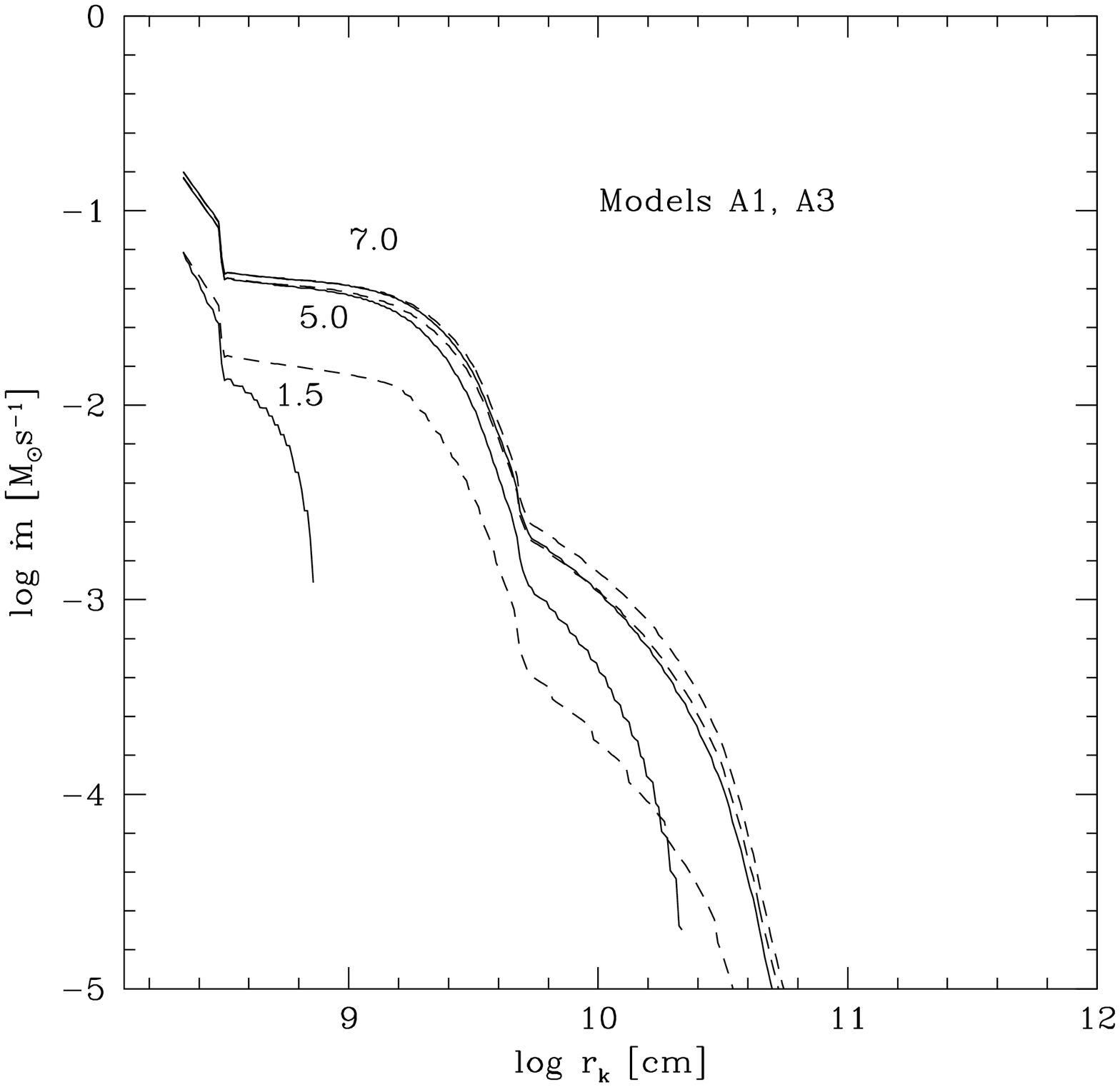}{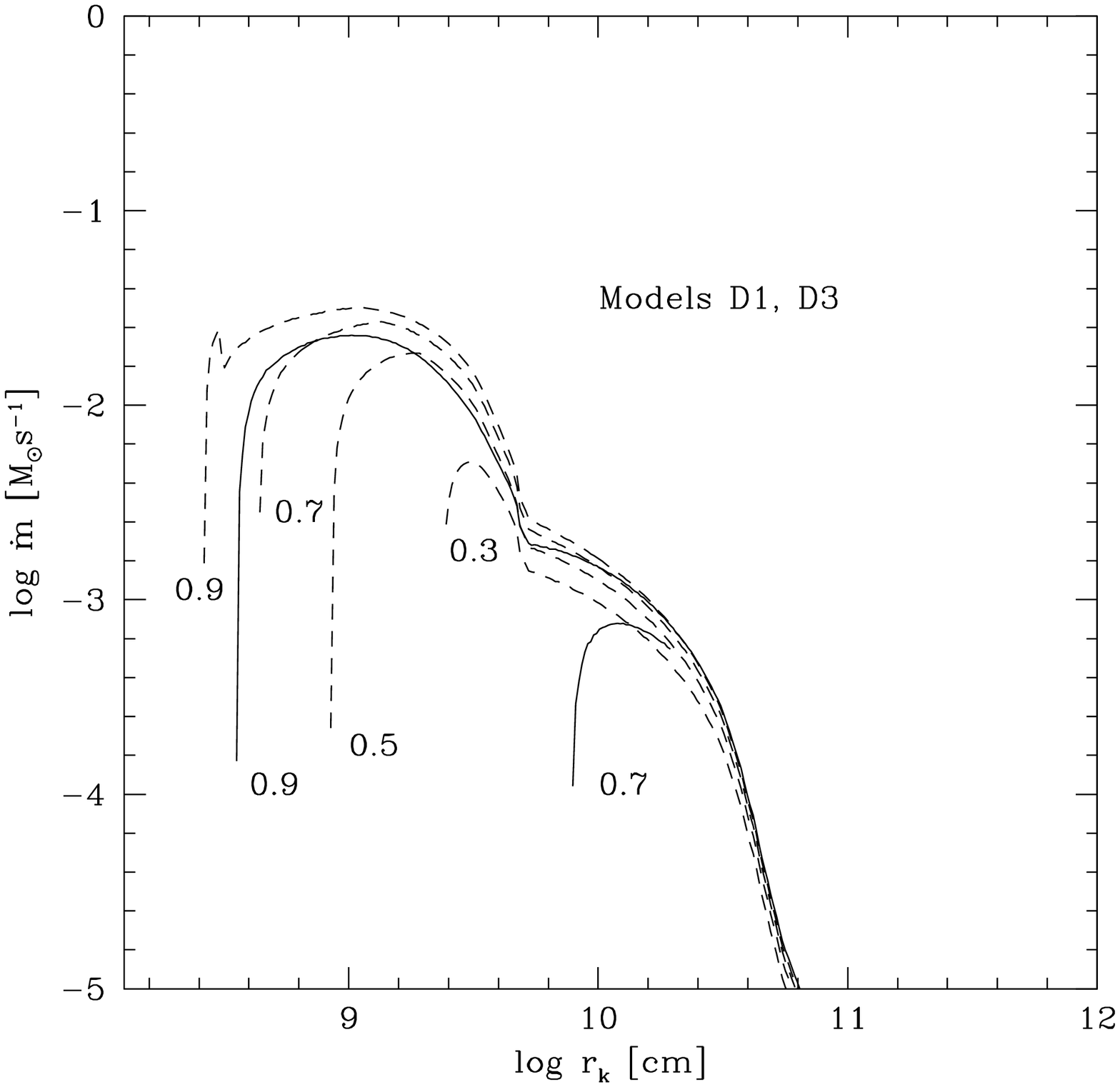}
\caption{ The instantaneous mass accretion rate during the collapse, i.e.
  as a function of
  radius $r_{\rm k}$ (the current inner radius of the envelope as it keeps
  accreting onto BH). 
  The plots show 2 models and 3 exemplary values of the normalization
  parameter $x$.  Left: $x=$7.0, 5.0 
  and 1.5 (marked by numbers); solid lines mark the accretion scenario (1)
  and dashed lines mark scenario (3).  Right: $x=$0.7, 0.5
  and 0.05 (marked by numbers); scenarios (1)
  - solid lines; scenario (3) - dashed lines.}
\label{fig:mdotinst}
\end{figure}

The Figure \ref{fig:mdotinst} shows the instantaneous accretion rate
during the collapse, i.e. as a function of the current inner radius of the 
envelope, as it keeps accreting onto the BH.  
In models {\bf A1}, {\bf A2} and {\bf A3},
the torus exists from the very beginning of the collapse, and
then the accretion rate is the largest, 
 equal about 0.08-0.15 $M_{\odot}$ s$^{-1}$, depending
on $x$.  Later, as the outer shells accrete, the accretion
rate drops and for all the models it is less than 0.01 $M_{\odot}$
s$^{-1}$ at $\log r$ = 8.8-9.8, depending on $x$.

In models {\bf D1}, {\bf D2} and {\bf D3}, 
the torus formation is delayed, because the faster rotating 
shells are in the outer parts of the 
envelope. For large $x$, the torus is formed already for 
log$r \sim$ 8.4-9.0, where the free-fall time scale is very short and the 
accretion rate is large, on the order of  0.02-0.04 $M_{\odot}$ s$^{-1}$.
For small $x$ (i.e. $x < 0.7$ for model {\bf D1} and  $x < 0.3$
for model  {\bf D3}), 
the torus does not form until the outermost shells accrete, and therefore
the maximum accretion rates obtained in these models are always below
0.01 $M_{\odot}$ s$^{-1}$.

This will have important implications, because 
as shown in a number of studies of the hyperaccreting tori in the GRB
central engine, for the accretion rates smaller than about 0.01 
$M_{\odot}$ s$^{-1}$ 
 the neutrino cooling becomes inefficient (see e.g. Popham, Woosley \& Fryer 1999;
Di Matteo et al. 2002; Janiuk et al. 2004).  
Therefore it is reasonable to limit our
definition of an 'active' central engine to such a minimum accretion
rate.

\begin{figure}
\epsscale{.80}
\plottwo{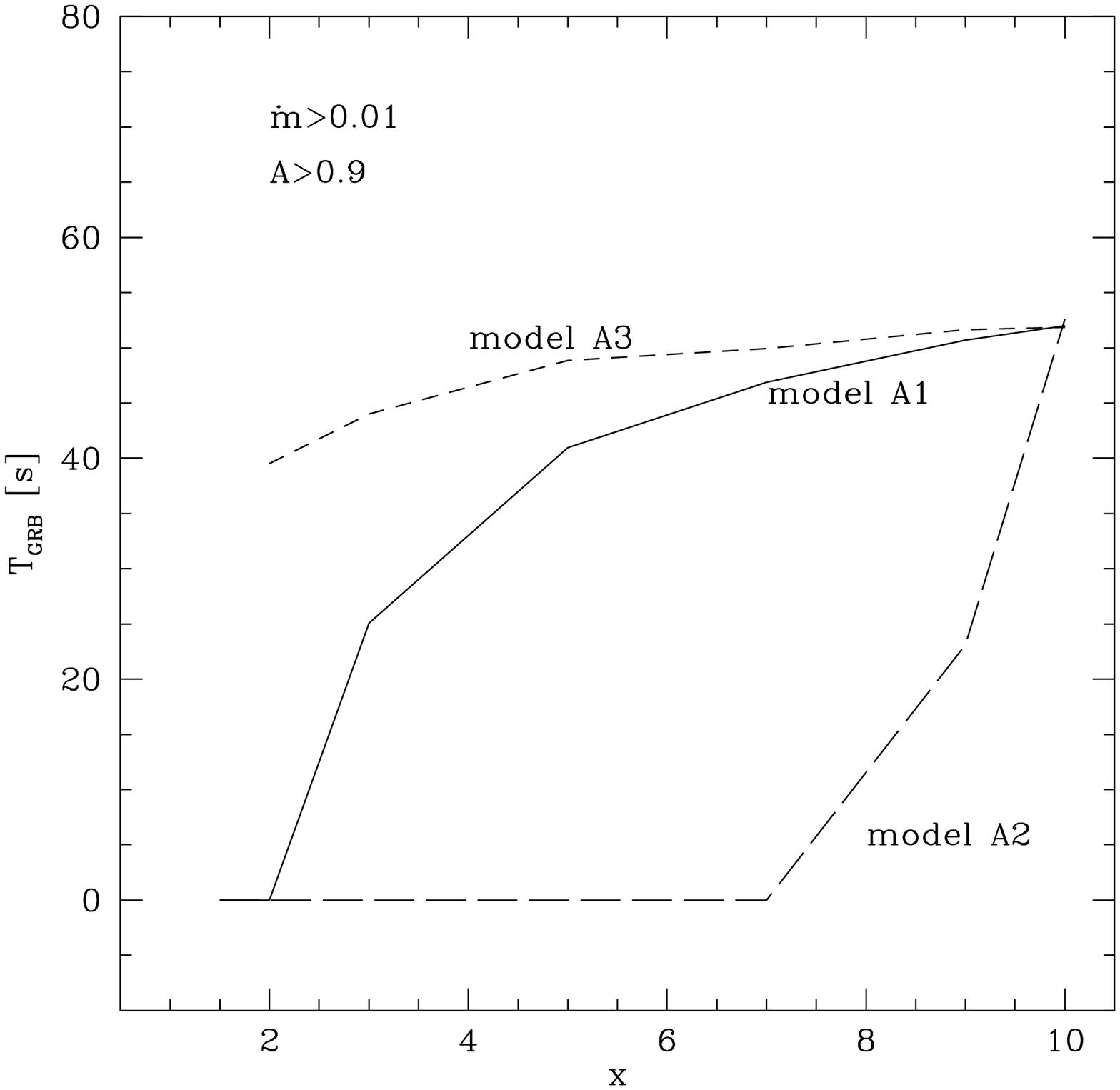}{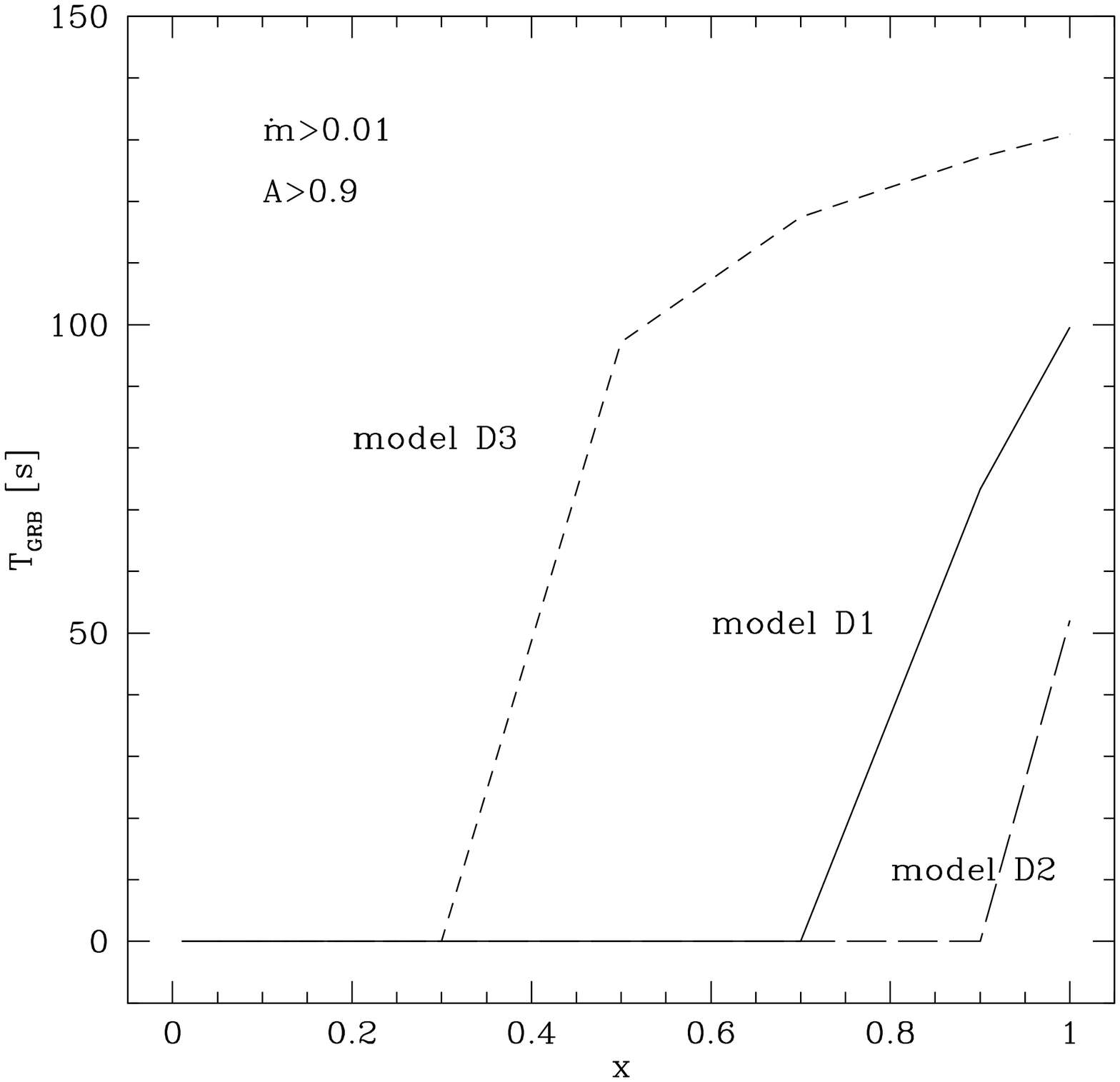}
\caption{The total duration of a GRB.  Left panel: models {\bf A}
  for the specific angular momentum distribution; Right panel: models {\bf D}. 
  The three accretion scenarios are
  uniform accretion (A1, D1; solid lines), polar and then torus accretion 
  (A2, D2; long dashed lines) and torus accretion (A3, D3; short dashed
  lines). The plots are in the function of the initial normalization of the specific
  angular momentum.  The calculated time  $T_{\rm GRB}$ results
  from the assumption of a minimum accretion rate, $\dot m_{\rm
    min}=0.01 M_{\odot}$s$^{-1}$, and the minimum 
  BH spin of $A_{\rm min}=0.9$. This time refers to the 'early' jet; see Sec. \ref{sec:diss}.}
\label{fig:dur}
\end{figure}

Another limitation for an efficient central engine will be the minimum
spin of the BH. 
Here we assume a conservative value of $A_{\rm
  min}=0.9$, to provide the energy source for the jet
(McKinney 2005).
The Figure \ref{fig:dur} shows
the duration of the central engine activity as a function of $x$, for
both models {\bf A} and {\bf D} and for all the 3 accretion scenarios. 
The plots account for the central engine activity time, when both assumptions
are satisfied, i.e. the accretion rate must be larger 
than $\dot m_{\rm min}$ and the BH spin must be larger than $A_{\rm min}$.

As the Figure shows, 
the torus accretion
scenario, marked with a dashed line,
leads to the longest duration of a GRB: up to 50 seconds in model {\bf A3}
and up to 130 seconds in model {\bf D3}.
 In this scenario, the BH spin is always
larger than our minimum value, and in most cases the BH
was spun up to $A_{\rm Kerr}=0.9999$. Therefore in practice, what determines
the GRB duration in this case, is the accretion rate. Consequently, the model
{\bf A3}, in which the accretion rate is larger, results in shorter GRBs 
than the model {\bf D3}.

The uniform accretion scenario
leads to a shorter GRB than in case of a torus accretion. Now, also the 
condition for a minimum BH spin is important, because for some models
the BH has not spun up or has been spun down, below $A_{\rm Kerr}=0.9$.
In the model {\bf A1}, for very small $x$, no GRB was produced.
Moreover, in model {\bf D1} the GRBs occurred only for $x>0.7$ (neither the spin 
nor the accretion rate condition was satisfied for smaller $x$), and the longest 
GRB duration  was $T \approx 100$ s.

The shortest GRBs were produced by scenario (2), i.e. the two
 phase accretion (obviously,
only in these models which had the second phase with a torus accretion). 
In models {\bf A2} and {\bf D2}, the activity of the GRB central engine 
was never longer than 50 seconds.


  The Figure \ref{fig:dur} shows the duration of a GRB resulting
  from the assumption of a minimum accretion rate {\it and} the minimum 
  BH spin.
  However, as we discuss below in Section \ref{sec:diss}, these two conditions refer
  to the two various mechanisms of powering the jet, which is emitting the gamma rays.
  The time $T_{\rm GRB}$ is different, if we consider only one of these mechanisms, i.e. impose only one of the above conditions.
 For instance, if we took into account only the condition for a minimum 
 BH spin, in model {\bf A3}  the GRB was 
 up to 4 times longer than that presented in the left panel of Fig. \ref{fig:dur}.
 Also, the model {\bf D3}
 produced long GRBs powered only by the BH spin, for
 which the minimum accretion rate condition was not satisfied (the models with $x<0.3$).
 The very long GRB durations for the spin
 condition result from the fact that at the end 
of the collapse, the BH is still spinning fast, while the accretion rate is 
very small and the mass accreted through the torus is large. 
 On the other hand, taking into account only the condition for the accretion rate, regardless of the BH spin, led to somewhat shorter (sometimes even two times shorter) 
GRBs than these presented in Fig. \ref{fig:dur}.  
This comes from the fact that the 
largest accretion rate,
leading to a shorter GRB,
is always at the beginning of the collapse, when the BH has not yet spun up enough.

\section{Discussion and conclusions}
\label{sec:diss}

In this article, we studied the collapsar model for long GRBs, powered
by accretion onto a spinning BH, which formed from the core of a massive,
rotating Wolf-Rayet star. To describe the rotation of the stellar interior, we adopted 
two different analytical functions, accounting for either a differential rotation
(models {\bf A1}, {\bf A2}, {\bf A3}), or a constant ratio between the gravitational and centrifugal forces 
(models {\bf D1}, {\bf D2}, {\bf D3}). This study is an important test
 for the rotation models of the GRB progenitor stars
 (e.g. Heger et al. 2005; Yoon et al. 2006; Detmers et al. 2008).

To describe how the accretion proceeds during the collapse, we
adopted three different scenarios: (1) uniform accretion, (2) two phase accretion, 
first from the poles and then from the torus and (3) only torus accretion.
The accretion onto the BH is in our approach a homologous process, in which
 the subsequent shells of the envelope 
add their mass to the central object. The angular momentum is also accreted,
but the limit for it is the critical angular momentum, to prevent the BH from
spinning with $A_{\rm Kerr} \ge 1.0$.
In this sense, we assume that the whole angular momentum 
with $l > l_{\rm crit}$, i.e. in the torus, 
is transported outwards. We do not invoke any particular mechanism of 
transport (i.e. the viscosity), and the momentum is taken out by a 
negligibly small amount of mass (e.g., Pringle 1981).
This simplified approach describes well a 
more realistic situation, in which the matter with small and large angular momentum
can be mixed. Therefore some parts of the gas with large $l_{\rm spec}$ might reach 
the BH,
while some other parts might be blown out with the polar outflow.

We focused on the evolution of the BH spin during the collapse. The large BH spin
is important for GRB production in two ways: 
first, to power the jet emission via the Blandford-Znajek (BZ) mechanism, and second,
because it alters the condition for the torus formation, i.e. 
the critical specific angular momentum.
We found that the spin of the BH strongly depends on both the model of the
$l_{\rm spec}$ distribution and on the accretion scenario.

In the torus accretion 
(i.e. either the second phase of the scenario 2, models {\bf A2} and {\bf D2}, 
or the scenario 3,  models {\bf A3} and {\bf D3}), 
the accreting material has  specific angular momentum
always $l_{\rm spec} \ge l_{\rm crit}$. This angular momentum must be transported 
outwards before reaching the BH, so that the gas which is changing the BH spin has
the specific angular momentum 
equal to $l_{\rm crit}$. Nevertheless, it is enough to spin up
the BH to the maximal rotation, $A_{\rm Kerr}=0.9999$, 
which happens in most cases at the very beginning of the collapse.
The polar accretion, i.e. the first phase of scenario 2. (models {\bf A2} and {\bf D2}),
leads only to the
BH spin-down in all the models.
The uniform accretion scenario is the most complex. 
In the model {\bf A1} it leads only to a temporary increase of the 
BH spin, while during the accretion of the outer shells, the BH is spinning down.
In the model {\bf D1}, the BH spin first decreases, while later during the collapse it 
may increase, provided the stellar envelope contains enough $l_{\rm spec}$.

We found that in model {\bf A1}, 
the final BH spin after the collapse is always about $A_{\rm end} \sim 0.85$, 
and it does not depend
on the normalization the specific angular momentum contained in the stellar envelope,
i.e. on $x$. However, 
the pattern of the BH spin evolution is very sensitive to this parameter.
Therefore for small values of $x$ it may happen that even for a short time during the collapse, the BH never reaches a spin 
$A_{\rm Kerr}>0.9$, 
 which we consider necessary to power 
the jet with the BZ mechanism. However, in the same models, the torus does exist
and the accretion rate in the torus is
large enough to power the jet via the neutrino annihilation.
This might lead to a relatively short living (less than $\sim$ 7-8 s) GRB 
central engine without a very rapidly spinning BH.
On the other hand, for $A_{\rm Kerr}>0.9$, the stage of a rapidly spinning BH
begins very shortly after the collapse has started, and
lasts much longer after the accretion rate in the torus has dropped 
below $\dot m=0.01$ M$_{\odot}$s$^{-1}$.
 For instance, a GRB powered by the BZ mechanism
may last almost $\sim 120$ s, while that powered
by the neutrino annihilation (concurrent with the spinning BH) lasted only $\sim 40$ s.
A very short time required for the BH to spin up, while the collapse proceeds, is
of the order of $\sim 1.5$ s.

In model {\bf D1} the situation is different. 
Here we do not find any models with only the 
neutrino-powered bursts, i.e. with a large accretion rate but 
not accompanied with a rapidly spinning BH.
In other words, whenever there exists a 
torus with a large accretion rate, the BH is spun up to $A_{\rm Kerr}>0.9$, and
 the timescale for this spin up is a fraction of a second ($\sim 0.15$ s).
Similarly to model {\bf A1}, the stage of a large BH spin can last much longer,
after the accretion rate has dropped below  $\dot m=0.01$ M$_{\odot}$s$^{-1}$. 
For instance, the BZ-powered burst lasting $\sim 430$ s is accompanied 
by a $\sim 100$ s burst powered by both BZ and neutrino mechanisms.

Observationally, this behavior may have led to
three kinds of jets. The first is a very short, lasting between a 
fraction of a second and few seconds, 'precursor' jet,
powered by only the neutrino annihilation, before the BH spins up. 
The second is
an 'early' jet, lasting several tens of seconds and 
powered by both neutrino and BZ mechanisms. The third is a 'late' jet, 
powered by only the spinning BH via the BZ mechanism.
In our models, we can have the GRB jets with all the three components, as well as
the 'orphan precursor' jets, when the BH failed to spin up.

The precursors have been detected
by Ginga, BeppoSAX, BATSE, INTEGRAL and Swift in some GRBs 
(e.g. Murakami et al. 1991; 
Piro et al. 2005; Lazzati 2005; Romano et al. 2006; McBreen et al. 2006). 
These GRB precursors are an important observational test for their
theoretical models (e.g. Ramirez-Ruiz et al. 2002;
Umeda et al. 2005; Morsony et al. 2007;  Wang \& Meszaros 2008).
For instance, in the sample of BATSE bursts, studied by Lazzati (2005), about 20\% 
of the bursts had a precursor, which was characterized by a 
 non-thermal spectrum and contained less than 1 per cent of the total counts.
The main GRB in these events was 
delayed with respect to the precursor by 10-200 seconds. 
As argued by Morsony et al. (2007), who 
in the 2-D numerical MHD simulations
identified three distinct phases during the jet
 propagation, this large gap in the emission might be a selection effect. 
Because of different opening angles of these three jets,
some observers located at large viewing angles may see a 'dead' phase, i.e. the break in the emission, related to the
second jet. 
Another explanation of the gap  between the
precursor and the main jet could be the 
development of the instabilities in the hyper-accreting disk (Perna et al. 2006; 
Janiuk et al. 2007), possibly combined with the viewing angle effects.

We therefore conclude that in the present model, the 'dead' phase would  refer to an 
'early' jet, which is powered by both neutrino and BZ mechanisms and can be
collimated to a much narrower angle than the 'late' jet. For the viewing angle larger 
than  the 'early' jet but smaller than the 'late' jet opening angle,
the observer should see the precursor, followed by a gap in the emission 
on the order of 40-150 seconds, and then see the 'main' GRB.
We also notice that recently, 
the observation of the bright, long GRB 080319B (Racusin et al. 2008),
seems to have confirmed that the jet's opening angle may vary, indicating for the two types of jets.

Finally, 
comparing our current models with the results presented for a non-rotating BH  
(Paper I), we 
notice that the GRB durations are similar in case of model {\bf A1}, i.e. 
$\sim 40$ s vs. $\sim 50$ for the Schwarzschild and Kerr BH main jet, respectively.
In model {\bf D1}, the discrepancy is more pronounced, namely
$\sim 50$ s vs. $\sim 100$ s, respectively. On the other hand, in the current 
work, the model  {\bf D1} produces GRBs powered by neutrino annihilation
only for a very narrow range of parameters (i.e. $x$),
while in Paper I for this model we found no limitations for $x$.

\acknowledgements
This work was supported in part by 
grant 1P03D00829 of the Polish State Committee for Scientific
Research, the Polish Astroparticle Network
grant 621/E-78/SN-0068/2007 and NASA under grant
NNG05GB68G.

\appendix

\section{Spin evolution}
\label{sec:rafal}
Space-time around astrophysical BHs is described by the Kerr
metric with two parameters, the mass-energy $M$, and the angular
momentum $J$.  Below we use dimensionless angular momentum $A \equiv
cJ/GM^2$ and dimensionless radius $\bar r = rc^2/GM$.

\noindent
The change of BH parameters due to accretion of rest mass
${\rm d}m$ is given by (Moderski \& Sikora 1996; see also Moderski et al. 1998):
\begin{eqnarray}
c^2 {\rm d} M = e {\rm d}m \,,\label{equ:evm}\\
{\rm d} J = l {\rm d}m \,,\label{equ:evj}
\end{eqnarray}
where $e$ and $l$ are the specific energy and angular momentum of
accreted matter, respectively.

\noindent
Combination of equations (\ref{equ:evm}) and (\ref{equ:evj}) gives the
evolution equation for the BH spin:
\begin{equation}
\frac{{\rm d} A}{{\rm d} \ln M} = \frac{\bar l}{\bar e} - 2A \,,
\label{equ:evA}
\end{equation}
where dimensionless quantities $\bar l \equiv cl/GM$ and $\bar e \equiv
e/c^2$ can be functions of $A$.  Thus whether spin increases or
decreases depends on the sign of the expression $\bar j - 2A \bar e$.

\subsection{Geometrically thick disk}
We will consider an accretion from geometrically thick disc as an
example.  In such a case the inner edge of the accretion disc is
located at the marginally bound orbit, $r_{mb}$ and
\begin{equation}
\bar r_{mb} = 2 \mp A + 2(1 \mp A)^{1/2} ~~~~ A = \pm \bar
r_{mb}^{1/2} (2 - \bar r_{mb}^{1/2})
\label{equ:rmb}
\end{equation}
\begin{equation}
\bar e_{mb} = 1
\end{equation}
\begin{equation}
\bar l_{mb} = 2 \bar r_{mb}^{1/2}
\label{eq:lmb}
\end{equation}
where upper signs are for direct accretion, while lower signs are for
retrograde accretion, respectively.

\noindent
The solution of equation~(\ref{equ:evA}) is (Abramowicz \&
Lasota~1980):
\begin{equation}
\bar r_{mb} M^2 = \bar r_{mb0} M_0^2 \,,
\label{equ:sol}
\end{equation}
where $\bar r_{mb0}$ is the initial marginally bound orbit.

\noindent
From equations (\ref{equ:sol}) and (\ref{equ:rmb}) we obtain
\begin{equation}
A = \cases {\pm {\bar r_{mb0} M_0 \over M} {(2 - {\bar r_{mb0} M_0
\over M})} & {\rm for} ${M\over M_0} \le \sqrt{\bar r_{mb0}}$ \cr \pm
1 & {\rm for} ${M\over M_0} \ge \sqrt{\bar r_{mb0}}$ \cr},
\label{equ:solA}
\end{equation}
where, for a given $A_0$, the value of $\bar r_{mb0}$ can be
calculated from equation (\ref{equ:rmb}).

\noindent
Formula (\ref{equ:solA}) together with the solution of
equation~(\ref{equ:evm})
\begin{equation}
M = M_0 + m
\end{equation}
describe the evolution of $A$ as a function of the amount of the accreted rest 
mass $m$.

\end{document}